\documentclass[useAMS,usenatbib]{mn2e}
\usepackage[T1]{fontenc}
\usepackage{aecompl}
\usepackage{graphicx}
\usepackage{latexsym}
\usepackage{psfig}
\usepackage{amssymb,amsmath}
\usepackage{subfig}
\usepackage{caption}

\title{Before the first supernova: combined effects of HII regions and winds on molecular clouds}

\author[J. E. Dale, J. Ngoumou, B. Ercolano, I.A. Bonnell]{J. E. Dale$^{1,2}$\thanks{E-mail: dale@usm.lmu.de (JED)}, J. Ngoumou$^{2}$, B. Ercolano$^{1,2}$, I. A. Bonnell$^{3}$\\
$^{1}$Excellence Cluster `Universe', Boltzmannstr. 2, 85748 Garching, Germany.\\
$^{2}$Universit\"{a}ts Sternwarte M\"{u}nchen, Scheinerstr. 1, 81679 M\"{u}nchen, Germany.\\
$^{3}$Department of Physics and Astronomy, University of St Andrews, North Haugh, St Andrews, Fife KY16 9SS}

\begin{document}

\pagerange{\pageref{firstpage}--\pageref{lastpage}} \pubyear{2006}

\maketitle

%\label{firstpage}

\def\mnras{MNRAS}
\def\apj{ApJ}
\def\aj{AJ}
\def\aap{A\&A}
\def\apjl{ApJL}
\def\apjs{ApJS}
\def\araa{ARA\&A}
\def\pasp{PASP}
 
\begin{abstract}
We model the combined effects of photoionization and momentum--driven winds from O--stars on molecular clouds spanning a parameter space of initial conditions. The dynamical effects of the winds are very modest. However, in the lower--mass clouds, they influence the morphologies of the HII regions by creating 10pc--scale central cavities.\\
The inhomogeneous structures of the model GMCs make them highly permeable to photons, ionized gas and supernova ejecta, and the leaking of ionized gas in particular strongly affects their evolution, reducing the effectiveness of feedback. Nevertheless, feedback is able to expel large fractions of the mass of the lower escape--velocity clouds. Its impact on star formation is more modest, decreasing final star formation efficiencies by 10--20$\%$, and the rate of change of the star formation efficiency per freefall time by about one third. However, the clouds still form stars substantially faster than observed GMCs.
\end{abstract}

 
\section{Introduction}
Stellar feedback operates at every scale in the star formation process and is a crucial ingredient in models of galaxy formation and evolution. Radiation pressure, jets and outflows in the 100au--1pc regime may help to set stellar masses, and may drive small--scale turbulence \citep[e.g][]{2006ApJ...640L.187L,2011ApJ...740..107C}. Radiative heating by protostars, including the low--mass objects, is also important at these scales for its influence on the fragmentation of the gas \citep[e.g][]{2007ApJ...656..959K,2009MNRAS.392.1363B,2010ApJ...710.1343U,2010ApJ...713.1120K}. At intermediate scales, HII regions, winds and radiation pressure driven primarily by massive stars profoundly alter the structures and velocity fields of giant molecular clouds (GMCs), creating bubbles and champagne flows \citep[e.g][]{1979MNRAS.186...59W,1979A&A....71...59T}, expelling gas from clouds' potential wells \citep[e.g][]{2002ApJ...566..302M} and, in some locations, triggering star formation \citep[e.g][]{2009ApJ...694L..26G,2011arXiv1109.3478W}. At galactic scales, the photons, momentum and energy produced by these processes and by supernovae shape the structures of galaxies and the interstellar medium and play a major role in globally regulating star formation  \citep[e.g][]{2013arXiv1311.2073H,2013ApJ...770...25A}.\\
\indent All star formation occurs in GMCs, so all stellar feedback mechanisms interact initially with the host GMC of their originating stars. While supernovae are likely to be the most important form of feedback on galactic scales, their effects will be modulated by the environment in which they explode. This in turn is modified by the other types of feedback in the interval between the onset of star formation and the demise of the first O--stars. Disentangling the effects of the various types of stellar feedback is exceedingly difficult from either an observational or theoretical point of view. There has, however, been considerable progress, particularly on the 1--100pc scales where the dominant mechanisms are photoionization, radiation pressure and stellar winds.\\
\indent Several authors have examined the issue of the relative or combined effects of winds and photoionization. \cite{1984ApJ...278L.115M} and \cite{1987ApJ...317..190M} argue that wind bubbles are likely to be trapped by the HII regions (HIIRs) of their driving stars, or that the inhomogeneous density fields in clouds are likely to cause the winds to cool on their expansion timescales or shorter, decreasing their effectiveness. \cite{2001PASP..113..677C} concluded that, except in very dense ($>$10$^{5}$ cm$^{-3}$) gas, the effects of photoionization will dominate. Analytic work by \cite{2002ApJ...566..302M} also found that wind bubbles were likely to be confined by HIIRs except in the case of very luminous clusters/associations.\\
\indent Most numerical work has focussed on single feedback mechanisms and the most popular choice has been HIIRs \citep[e.g][]{2005MNRAS.358..291D,2010ApJ...719..831P,2010ApJ...715.1302V} or jets \citep[e.g][]{2006ApJ...640L.187L,2010ApJ...709...27W,2012ApJ...747...22H,2012ApJ...754...71K}, although there has also been interest in winds.\\
\begin{figure*}
\includegraphics[width=0.99\textwidth]{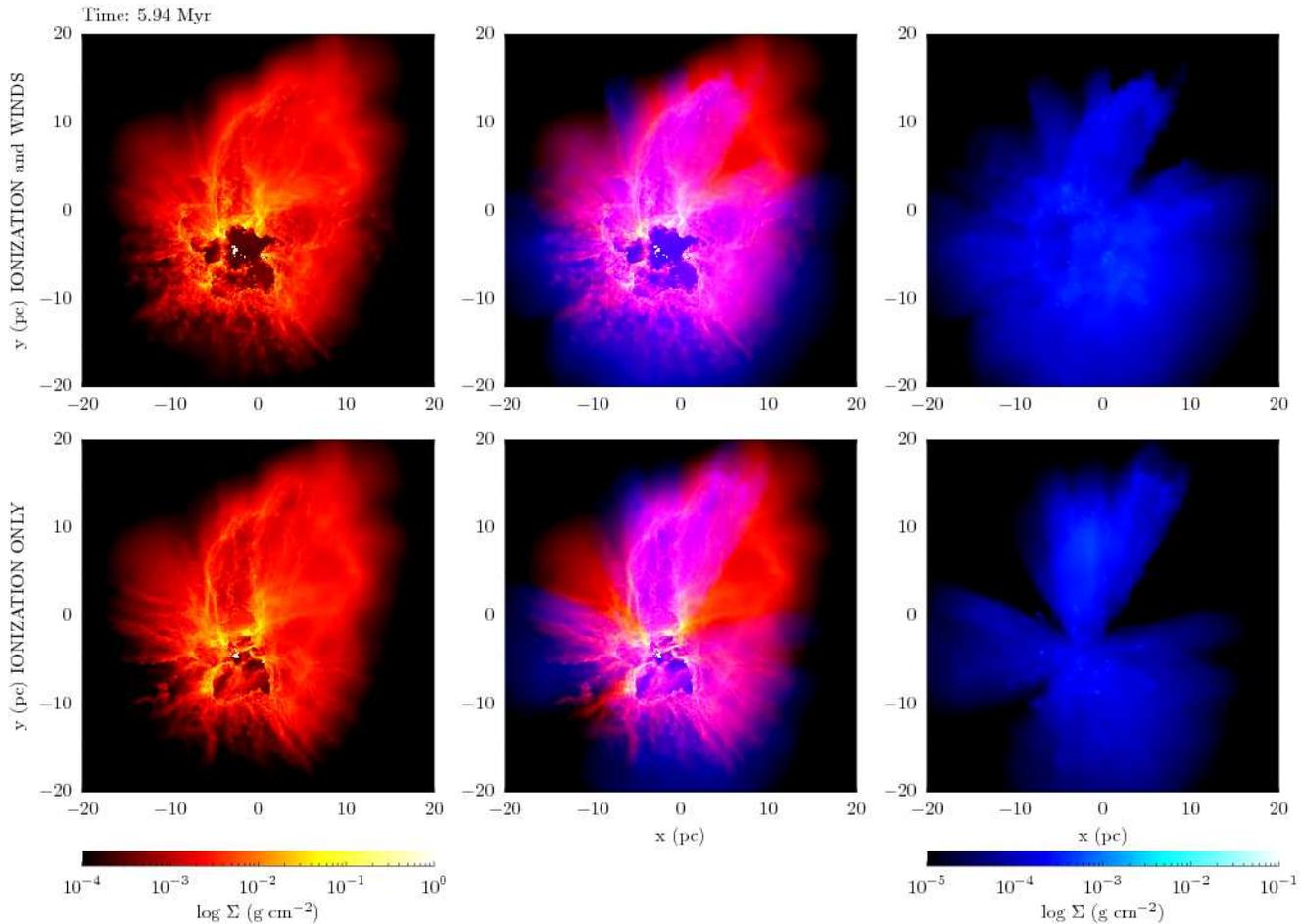}
\caption{Comparison of the cold gas (left panels) and the HIIRs (right panels) at an early epoch in the Run I calculations with (top row) and without (bottom row) winds included. Centre panels are superpositions of the left and right panels.}
\label{fig:hiir_splat}
\end{figure*}
\begin{figure*}
\includegraphics[width=0.99\textwidth]{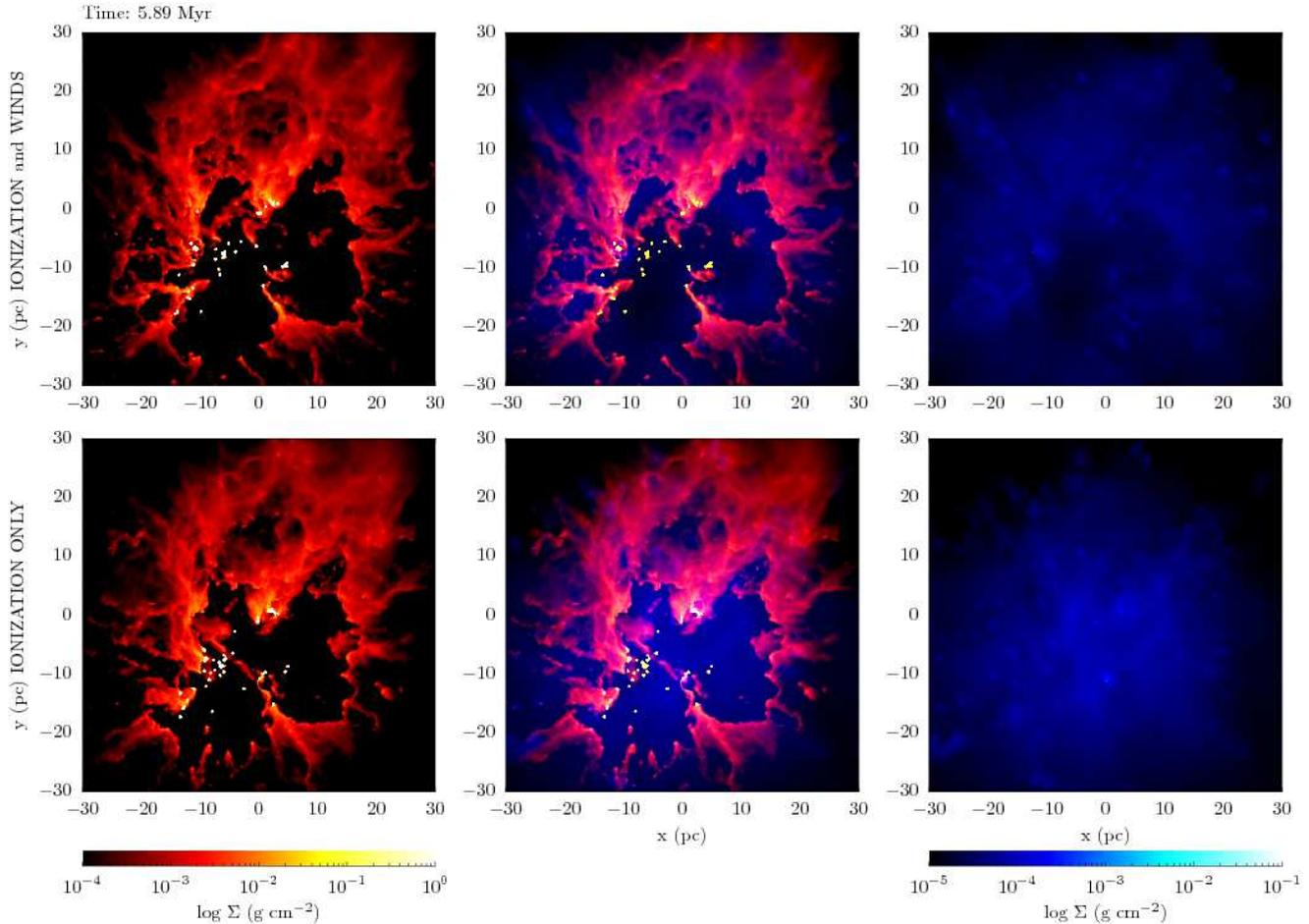}
\caption{Comparison of the cold gas (left panels) and the HIIRs (right panels) at the end of the Run UQ calculations with (top row) and without (bottom row) winds included. Centre panels are superpositions of the left and right panels.}
\label{fig:hiir_hole}
\end{figure*}
\indent \cite{2012ASPC..453...25F} and \cite{2013MNRAS.431.1337R} performed 3D simulations of winds blowing inside structured clouds and found that the winds were often able to carve escape routes out of the cold gas, venting out of the clouds and substantially reducing their destructive effects. \cite{2005MNRAS.358..291D} observed a similar effect operating within HIIRs in the absence of winds.\\
\indent \cite{2003ApJ...594..888F} and \cite{2006ApJ...638..262F} performed detailed 2D hydrodynamic simulations of the growth of wind bubbles inside HIIRs expanding into a smooth medium. They concluded that, for very massive stars ($>$60 M$_{\odot}$), the wind would sweep up the HIIR into a thin shell lining the inner wall of the feedback--driven bubble. However, for more modest stellar masses ($\sim35$ M$_{\odot}$), the wind bubble expansion stalled inside the ionized gas, resulting in an HIIR with a central hole, but otherwise having relatively little effect on the dynamical evolution of the bubble.\\
\indent Observational studies of this issue are inconclusive. \cite{2009ApJ...693.1696H} compared the one--dimensional wind--bubble model of \cite{1977ApJ...218..377W} to the Carina nebula. They found that the X--ray luminosity of the nebula is two orders of magnitude lower than the model predicts. This could be taken to mean either that the winds are not cooling, or that much of the hot wind gas has escaped from the bubbles. However, they also concluded that the high filling factor  of photoionized gas indicates that the dynamics of Carina are dominated by the HIIRs.\\
\indent \cite{2012ApJ...757..108Y} examined the effects of winds inside HIIRs on the ionization parameter, the ratio of ionizing photon number density to hydrogen number density. Windblown cavities inside HIIRs would decrease the ionization parameter by moving gas away from the stars, where the photon flux is largest. The ionization parameter can then be used to infer the degree of influence of winds on HIIRs. \cite{2012ApJ...757..108Y} could find no evidence for wind--dominated bubbles in their sample of Galactic and extragalactic HIIRs. They blame this on leakage of wind gas beyond the ionization front.\\
\indent \cite{2011ApJ...731...91L} studied the 30 Doradus region, comparing the radiation pressure, ionized gas pressure and X--ray emitting gas pressure as a function of radius from the R136 cluster. They concluded that the radiation momentum flux is larger than the ionized gas pressure within 75pc, that the reverse is true further out, and that the wind gas pressure is nowhere significant. \cite{2011ApJ...738...34P}, however, inferred that radiation pressure was largely unimportant in 30 Dor, but that the pressure of confined X--ray emitting wind gas has had substantial dynamical influence. \cite{2013arXiv1309.5421L} extend the work of \cite{2011ApJ...731...91L} to a sample of 32 HIIRs in the Magellanic Clouds. They report that the ionized gas pressure dominates the dynamics of the bubbles.\\
\indent In a series of recent papers, we have simulated the isolated effects of photoionizing radiation [\cite{2012MNRAS.424..377D}, \cite{2012MNRAS.427.2852D}, \cite{2013MNRAS.430..234D} and \cite{2013MNRAS.431.1062D}, hereafter Papers I--IV] \emph{or} stellar winds [\cite{2013arXiv1309.7355D}, hereafter Paper V] on a parameter--space of model turbulent GMCs. In all clouds, the winds acting alone did substantially less damage to the clouds than the HIIRs acting alone, and the degree of damage was a strong function of the cloud escape velocity. In this paper, we bring this work together by simulating the effects on the same model clouds of ionization \emph{and} winds.\\
\indent In Section 2, we briefly describe our numerical methods and summarize the properties of our model clouds. Section 3 contains the results of our simulations of turbulent clouds, and discussion and conclusions follow in Sections 5 and 6 respectively.\\
\section{Numerical methods}
\indent We perform Smoothed Particle Hydrodynamics (SPH) simulations of turbulent molecular clouds including the effects of photoionizing radiation and momentum--driven winds from massive stars.\\
\indent Our model clouds initially have a smooth spherical Gaussian density profile such that the central density is three times higher than that on the edge, which ensures that the clouds remain centrally condensed. The velocity field is initially turbulent with a Kolmogorov energy power spectrum $E(k)\propto k^{-5/3}$ giving an initial linewidth--size relation of $\Delta v\propto l^{\frac{1}{3}}$, populated in the wavenumber range 4--128. The total kinetic energy in the velocity field is scaled so that the virial ratio of the clouds is either 0.7 (`bound' clouds) or 2.3 (`unbound' clouds).\\
\indent Stars are represented by sink particles \citep{1995MNRAS.277..362B}. In our simulations of 10$^{5}$ and 10$^{6}$M$_{\odot}$ clouds, the sink particles represent stellar clusters, otherwise they represent individual stars. Where sink particles represent clusters, we assume for each sink a Salpeter mass function between $0.1$ and 100M$_{\odot}$ and compute how many stars more massive than 30M$_{\odot}$ the cluster would host. If this number is unity or larger, we multiply it by $2\times10^{48}$ s$^{-1}$ to obtain the total ionising flux of the cluster. We showed in the Appendix of Paper I that these assumptions agree well with the \emph{stellar} IMFs and ionising fluxes from the lower--mass clouds in which individual stars are modelled. The most massive stars in these runs are typically 30--40M$_{\odot}$ with one 65M$_{\odot}$ object in Run I. We also showed that uncertainties of a factor of a few in the ionising luminosities has negligible influence on our results.\\
\indent We combine the numerical methods from our previous work. We use the multisource ionization code described in \cite{2007MNRAS.382.1759D} in which Str\"omgren integrals are computed along rays to find the photon flux received at each particle from each source. When computing the path integrals for a given source, the recombination rates occurring in each particle are scaled by the fraction of the total photon flux received by that particle which comes from the given source, and the global solution is iterated on until the total ionisation fraction converges. This algorithm is described in more detail in Paper I.\\
\indent We also employ the multisource winds code described in \cite{2008MNRAS.391....2D} and Paper V. This algorithm provides a lower limit to the effects of stellar winds by modelling the momentum input from the winds only. Massive stars are regarded as the sources of spherically--symmetric momentum fluxes represented by non--hydrodynamic \emph{momentum packets}. The packets are emitted in random directions and the algorithm determines which SPH particle (if any) they strike, and applies an acceleration to that particle accordingly. No thermal energy is transferred. Most of the SPH particles struck by the wind in these simulations are ionised, and their temperatures are maintained at 10$^{4}$K.\\
\indent Our initial conditions are the same set of bound and unbound turbulent clouds described in Papers I and III, evolved from the same points in time when each had formed a few massive stars or massive subclusters. In Table \ref{tab:init}, we summarize the important parameters of all sixteen simulations.\\
\begin{table*}
\begin{tabular}{|l|l|l|l|l|l|l|l|l|l|}
Run&Mass (M$_{\odot}$)&R$_{0}$(pc)&$\langle n(H_{2})\rangle$ (cm$^{-3}$) & v$_{\rm RMS,0}$(km s$^{-1}$)&v$_{\rm RMS,i}$(km s$^{-1}$)&v$_{\rm esc,i}$(km s$^{-1}$)&$t_{\rm i}$ (Myr) &t$_{\rm ff,0}$ (Myr)\\
\hline
A&$10^{6}$&180&2.9&5.0&3.6&6.1&20.83&19.6\\
\hline
B&$10^{6}$&95&16&6.9&5.1&8.4&7.83&7.50\\
\hline
X&$10^{6}$&45&149&9.6&6.4&12.0&3.56&2.44\\
\hline
D&$10^{5}$&45&15&3.0&2.0&3.4&15.99&7.70\\
\hline
E&$10^{5}$&21&147&4.6&2.9&5.2&5.37&2.46\\
\hline
F&$10^{5}$&10&1439&6.7&4.2&7.6&2.24&0.81\\
\hline
I&$10^{4}$&10&136&2.1&1.4&2.3&5.37&2.56\\
\hline
J&$10^{4}$&5&1135&3.0&1.8&3.5&2.09&0.90\\
\hline
\hline
UZ&$10^{6}$&45&149&18.2&9.4&13.8&4.33&2.9\\
\hline
UB&$3\times10^{5}$&45&45&10.0&4.6&7.6&9.40&6.0\\
\hline
UC&$3\times10^{5}$&21&443&14.6&6.0&11.1&4.03&1.9\\
\hline
UV&$10^{5}$&21&148&12.2&3.9&6.4&10.44&3.3\\
\hline
UU&$10^{5}$&10&1371&8.4&5.8&9.3&3.73&1.1\\
\hline
UF&$3\times10^{4}$&10&410&6.7&3.5&5.1&3.28&2.0\\
\hline
UP&$10^{4}$&2.5&9096&7.6&3.6&5.9&1.83&0.4\\
\hline
UQ&$10^{4}$&5.0&1137&5.4&2.6&4.1&3.13&1.2\\
\hline
\end{tabular}
\caption{Initial properties of clouds listed in descending order by mass. Columns are the run name, cloud mass, initial radius, initial RMS turbulent velocity, RMS turbulent velocity at the time ionization becomes active, the escape velocity at the same epoch, the time at which ionization begins, and the initial cloud freefall time.}
\label{tab:init}
\end{table*}
\begin{figure}
\includegraphics[width=0.45\textwidth]{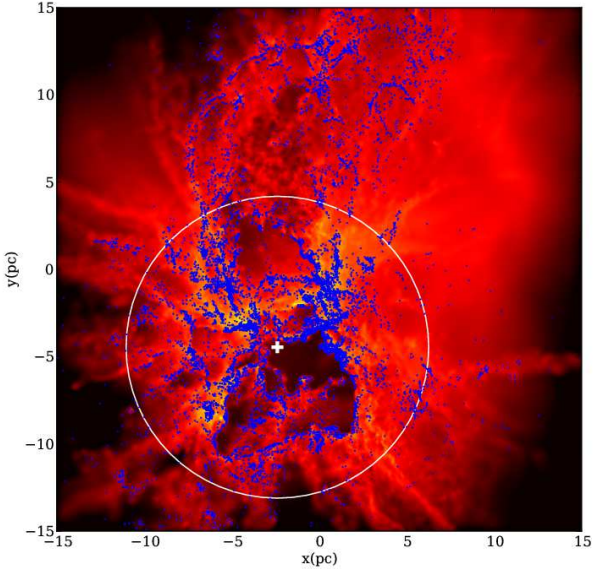}
\caption{Render from Run I with gas column density shown in red--orange--yellow colours, particles on the ionization fronts shown as blue dots, the location of the most massive ionizing source shown as a white cross, and the bubble defined by the mean distance of all ionization--front particles from the most massive source shown as a white circle.}
\label{fig:ifront_render}
\end{figure}
\begin{figure}
\includegraphics[width=0.45\textwidth]{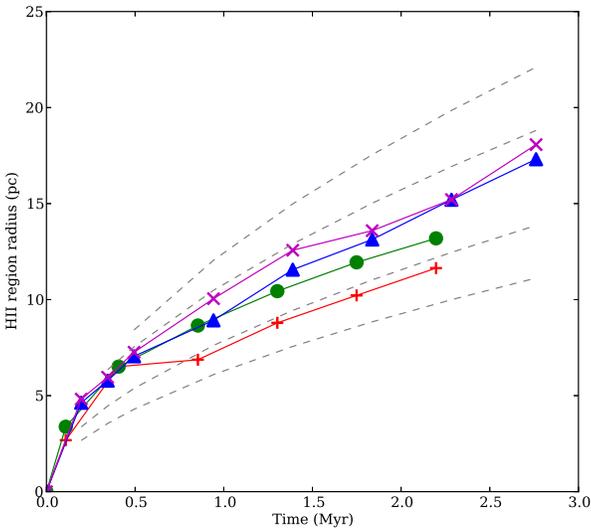}
\caption{Evolution of the ionization front radius in the Run I ionization--only (red), dual feedback (green) and Run UQ ionization--only (blue) and dual--feedback (magenta) calculations. Representative Spitzer solutions are shown as grey dashed curves.}
\label{fig:hiir_radius}
\end{figure}
\begin{figure}
\includegraphics[width=0.50\textwidth]{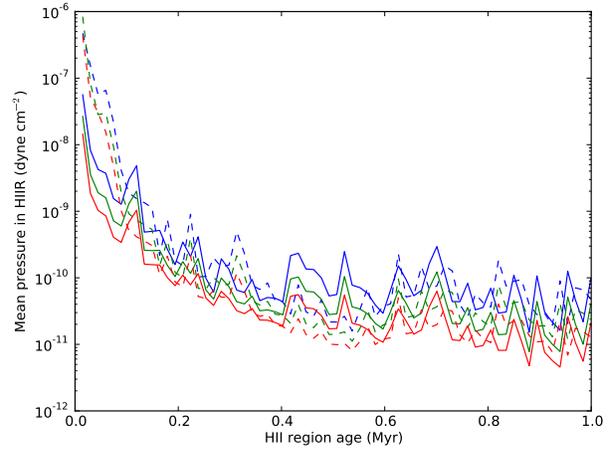}
\caption{Evolution of the pressure inside the 90 (red), 75 (green) and 50 (blue) percent of the ionized gas closest to the most massive stars in the ionization--only (solid lines) and dual--feedback (dashed lines) Run I calculations.}
\label{fig:hiir_press_runi}
\end{figure}
\begin{figure}
\includegraphics[width=0.45\textwidth]{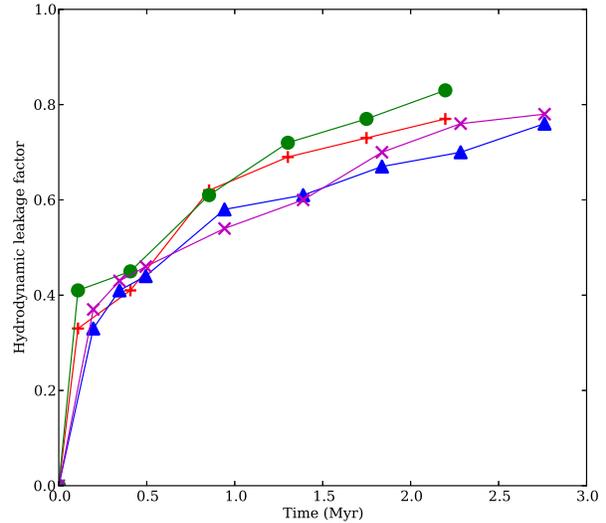}
\caption{Evolution of the gas leakage factors, the fractions of sky not covered by neutral gas at the ionisation front radius, in the Run I ionization--only (red), dual feedback (green) and Run UQ ionization--only (blue) and dual--feedback (magenta) calculations.}
\label{fig:hiir_hydro_leak}
\end{figure}
\begin{figure}
\includegraphics[width=0.45\textwidth]{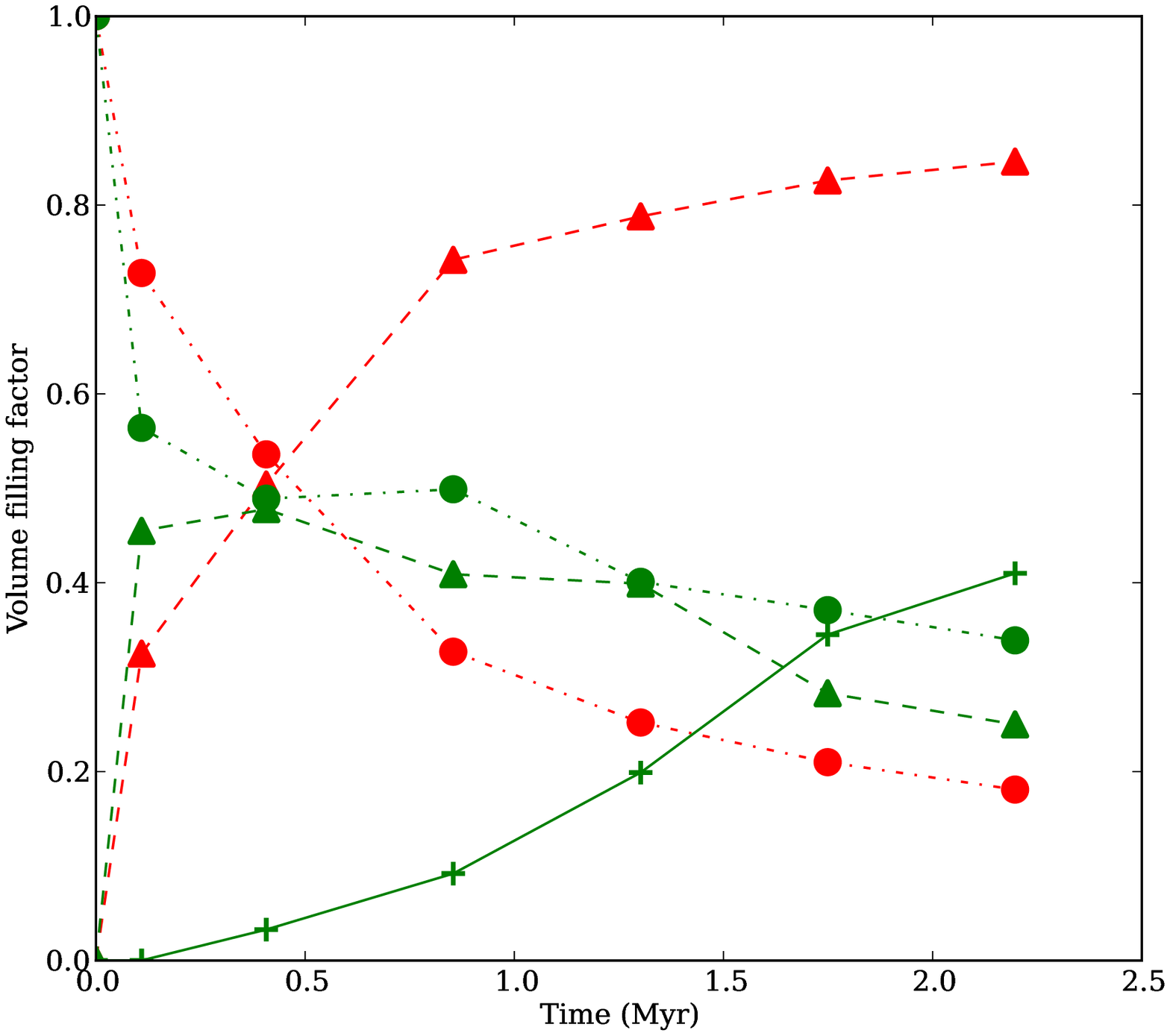}
\caption{Evolution of the volume filling factors of HII (triangles and dashed lines), wind (pluses and solid line) and cold neutral gas (circles and dot--dashed lines) within the ionization front radius in the ionization--only (red) and dual--feedback (green) Run I calculations.}
\label{fig:hiir_filling}
\end{figure}
\section{Results}
\subsection{HII region morphology}
In the lower--mass clouds, the winds modify the HIIR morphologies in two ways, depicted in Figures \ref{fig:hiir_splat} and \ref{fig:hiir_hole}. At early times, the winds help the HIIRs clear away the dense gas near the massive stars, as illustrated in Figure \ref{fig:hiir_splat}, from Run I. The evacuation of the cold gas (shown in the left panels) is more efficient in the dual--feedback run (top row) than in the ionization--only run (bottom row). This change is reflected in the shape of the HIIR. Dense material near the sources collimates the HIIR in the ionization--only calculation into conical lobes radiating from the central cluster. This morphology is not found in the seminal work of \cite{1989ApJS...69..831W} and is not a typical HIIR shape. However, the clearing away of the collimating material by winds results in a more circular projected morphology. The HIIR at this early epoch would probably be classified as core--halo, since it seems to be centrally--condensed, or multiply--peaked.\\
\indent Later, the winds create large ($\sim10$pc) holes inside the HIIRs, depicted in Figure \ref{fig:hiir_hole}. This shows the end of the Run UQ calculations, with the ionization--only run on the bottom and the dual--feedback run on top. The HIIR in the ionization--only simulation (lower--right panel) has an irregular morphology and would fall into the centrally--condensed or multiply--peaked categories of \cite{1989ApJS...69..831W}. The windblown HIIR (upper--right panel), by contrast, exhibits a hole whose shape is roughly the same as, but smaller than, the hole in the cold gas. The top central panel shows the HIIR as a thick lining of the inner surfaces of the bubbles. This HIIR would probably then be classified as shell--like by \cite{1989ApJS...69..831W}. Overall, the action of winds is to suppress the multiple--lobe morphological type which is not commonly seen, and to generate the shell--like form. From this perspective, the windblown HIIRs resemble observed systems better than those created by ionization alone.\\
\indent We examine the HIIR evolution in more detail by locating the ionization fronts and using their median separation from the brightest ionizing source to define a bubble radius and volume. Figure \ref{fig:ifront_render} shows the ionization--only Run I calculation 0.88Myr after feedback was enabled. A column--density projection of the gas is shown in red, orange and yellow. Neutral gas particles located just behind the ionization front(s) are shown as blue dots, the most massive star as a white cross and the median ionization front radius as a white circle. Note that the true shape of the ionization front is far from spherical.\\
\subsection{HII region expansion}
\indent Once the typical ionization front radius has been defined, we trace the expansion of the HIIR. Figure \ref{fig:hiir_radius} shows the time evolution of the median ionization front radius in the ionization--only and dual--feedback Runs I and UQ.\\
\indent The evolution of an HIIR in a uniform medium is described by the solution in \cite{1978ppim.book.....S}. If the temperature of the ionised gas is fixed, the solution describes a family of curves dependent only on the initial size of the ionised volume, the Str\"omgren radius, R$_{\rm s}$. $R_{\rm s}$ is not easy to estimate in inhomogenous gas, but we plot for comparison Spitzer evolution curves for $R_{\rm s}=$1.5, 2.5, 5.0 and 7.5 pc. This allows us to gauge qualitatively the effect of gas leakage and/or winds on the expansion of the HIIRs.\\
\indent The effect of either gas leakage or winds on the bubble expansion appears to be slight Neither process radically alters the expansion laws of the bubbles and the expansion curves do not differ markedly from the Spitzer solution for a closed bubble in a uniform cloud. These results are similar to those obtained by \cite{2012MNRAS.427..625W}, who modelled the influence of photoionization on fractal clouds. Their Figure 3 shows the evolution of the ionization front radius for values of the cloud fractal dimension D in the range 2.0--2.8. They found little change in the evolution with increasing fractal dimension, and the general form of the curves again deviates little from the Spitzer model.\\
\indent We examine the dynamical influence of the winds in Figure \ref{fig:hiir_press_runi}, which shows the pressure inside the ionisation--only (solid lines) and dual--feedback (dashed lines) HIIRs in Run I as functions of time. At very early times, the pressure in the windblown HIIR is roughly an order of magnitude larger, but when the HIIRs break out of the confining cold gas at $\sim10^{5}$ yr, the pressures rapidly become very similar.\\
\subsection{HII region gas leakage}
\indent Once we have defined the bubble radius using the median ionisation front, we divide the cloud into the `bubble', lying inside the ionization front, and the `cloud' which lies outside it. We first examine the leakiness of the bubbles to gas. We remove all gas outside them and the \emph{ionized} gas inside them and perform a Hammer projection from the point of view of the most massive star on the remainder. We then compute the fraction of sky \emph{not} covered by neutral gas and thus open to leakage of gas from the bubble.\\
\indent We show the evolution of these gas leakage factors in Runs I and UQ in Figure \ref{fig:hiir_hydro_leak}. The gas leakage factors increase with time as the cold gas is cleared away from the sources. The difference between the ionization--only and dual--feedback calculations is again slight. Gas leakage is small early in the simulations, but increases very rapidly over a few$\times10^{5}$ yr as the HIIRs burst out of the dense material in which the O--stars are born. At early times, gas leakage is somewhat larger in the dual--feedback simulations, as the winds initially aid the HIIRs in clearing gas from near the sources. Typical gas leakage factors for these two calculations are 0.6--0.7, whereas for the Run E cloud which is much less severely affected by feedback, the leakage factor is only 0.1--0.2.\\
\indent \cite{2009ApJ...693.1696H} constructed one--dimensional models of the Carina nebula in which they pointed out that the model of \cite{1977ApJ...218..377W} predicts a bubble too large for Carina's estimated age. They modified the standard model to include leakage of wind gas and found that the gas leakage fractions required to achieve agreement with the present state of Carina were $\approx0.5$, close to the values measured here (note that they discuss the covering factor $C_{\rm f}=1-f_{\rm leak}$).\\
\subsection{HII filling factors}
\indent We interpolated the SPH density field onto a grid and counted the fractions of cells inside the ionization front which are filled with HII or cold gas, or which have been cleared by winds. We show the time evolution of these filling factors in Run I in Figure \ref{fig:hiir_filling}. In the ionization--only calculation (red lines and symbols), the evolution is a trade--off between the neutral gas, whose filling factor starts high (as expected for a centrally--condensed cloud forming stars near its centre of mass) and falls rapidly to $<20\%$, and the HII which does the reverse. In a spherical HIIR, one would expect the volume behind the ionisation front to be entirely filled by ionised gas, so that the HII filling factor would be unity. That this is not so in our bubbles is a result of the bubbles not being spherical, so that some of the gas behind the median ionisation front is actually neutral. However, the bubbles clearly approach a more spherical form as time passes.\\
\indent In the dual--feedback run, the winds create a cavity inside the HIIR which grows rapidly over $\sim10^{5}$yr. The volume--filling factor of the cavity reaches $>40\%$ by the end of the calculation. This correspondingly reduces the filling factor of the ionized gas in the bubble, which rises to $\approx50\%$ before declining to $\approx20\%$. The filling factor of neutral gas in the bubble is higher in the dual--feedback calculation, implying that the bubble has a more complex shape in this calculation. Overall, the filling factors of ionized gas remain large -- $>10\%$ -- in reasonable agreement with the findings of \cite{2009ApJ...693.1696H}.\\
\subsection{Influence of feedback on cold gas morphology}
\begin{figure*}
\includegraphics[width=0.99\textwidth]{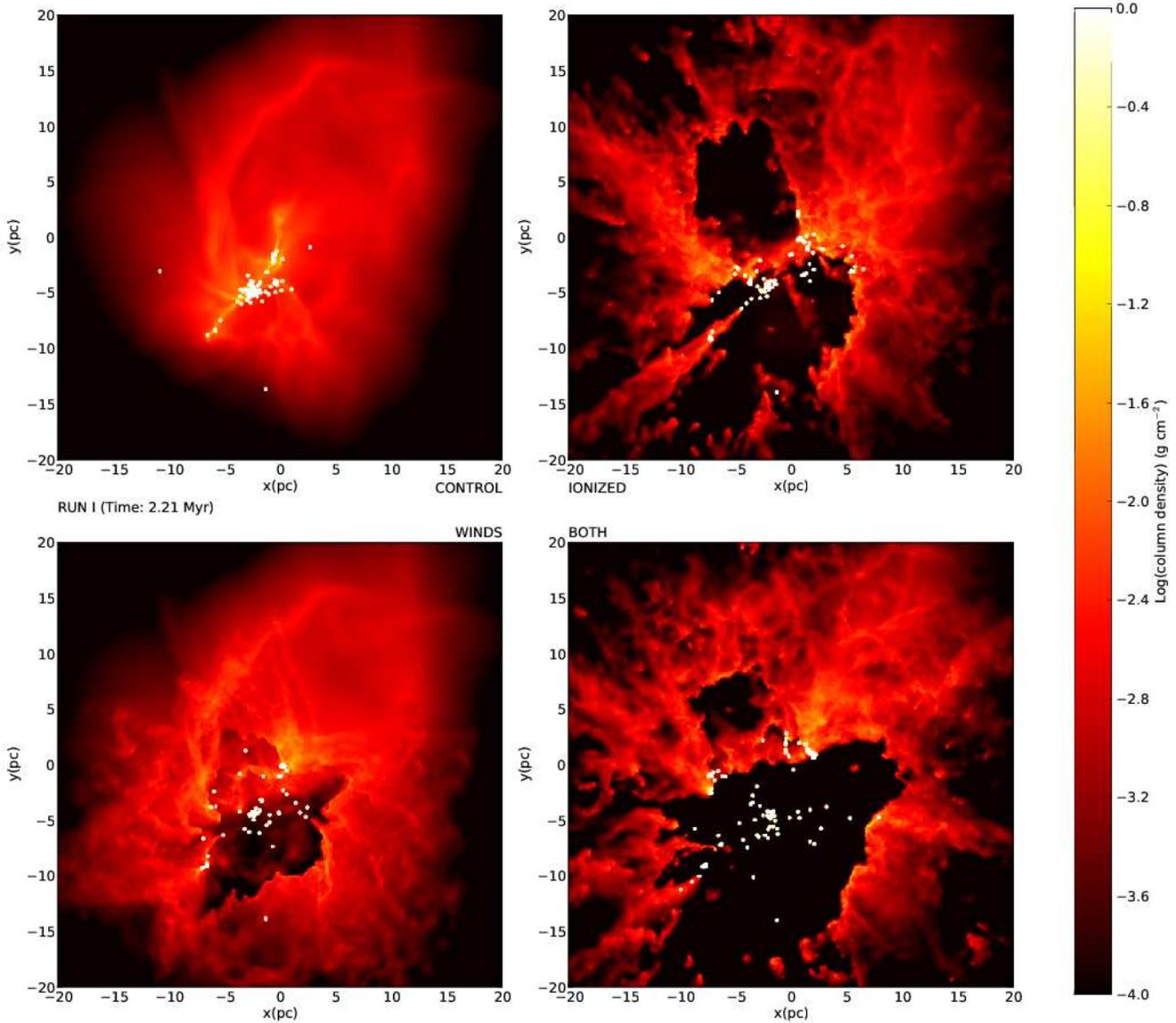}
\caption{Comparison of the cold gas morphology in the Run I calculations with no feedback (top left), with ionization only (top right), with winds only (bottom left) and with both winds and ionization (bottom right).}
\label{fig:runi_all}
\end{figure*}

\begin{figure*}
\includegraphics[width=0.99\textwidth]{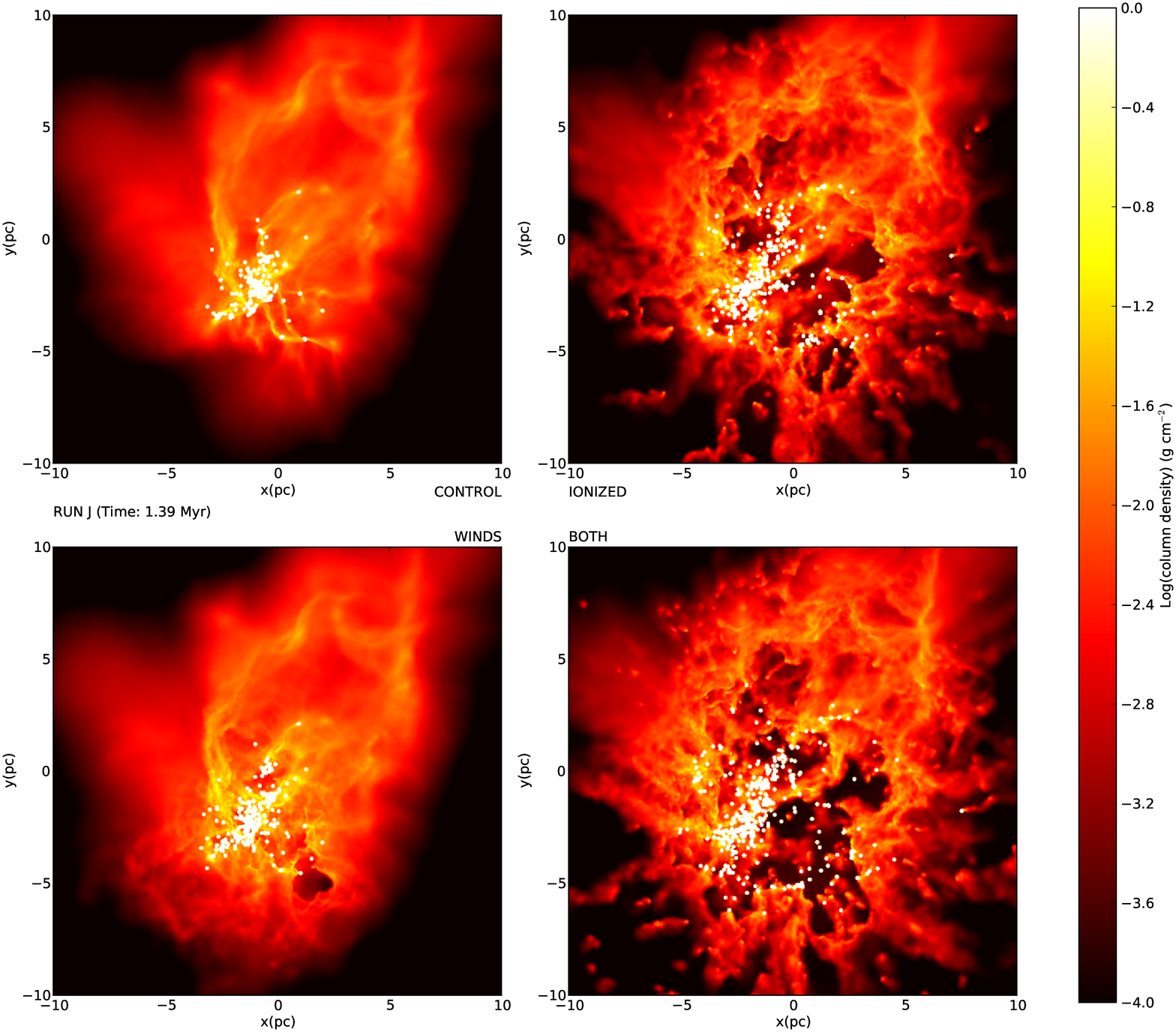}
\caption{Comparison of the cold gas morphology in the Run J calculations with no feedback (top left), with ionization only (top right), with winds only (bottom left) and with both winds and ionization (bottom right).}
\label{fig:runj_all}
\end{figure*}

\begin{figure*}
\includegraphics[width=0.99\textwidth]{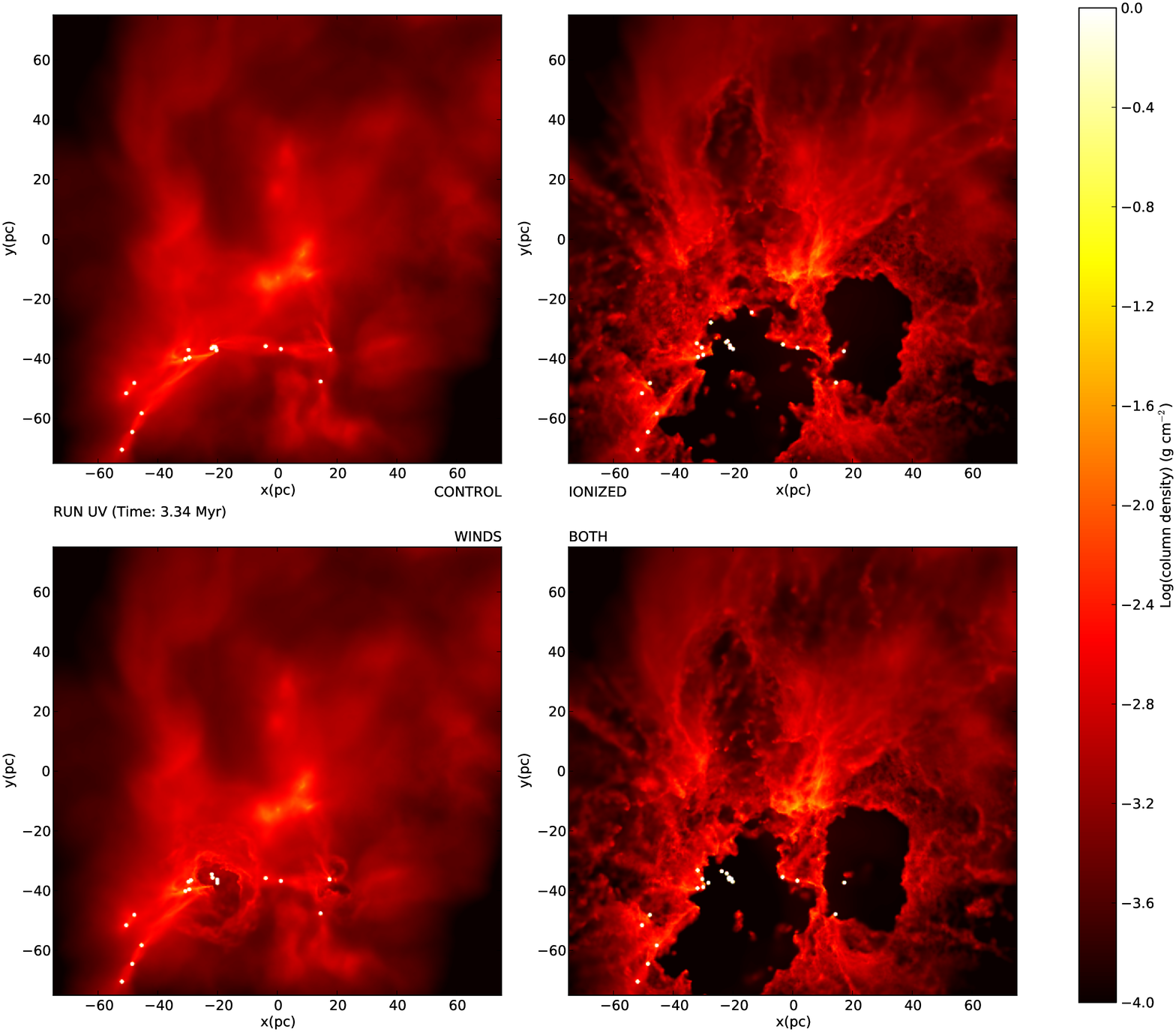}
\caption{Comparison of the cold gas morphology in the Run UV calculations with no feedback (top left), with ionization only (top right), with winds only (bottom left) and with both winds and ionization (bottom right).}
\label{fig:runuv_all}
\end{figure*}

\indent In Figures \ref{fig:runi_all}, \ref{fig:runj_all}, \ref{fig:runuv_all}, we compare the final cold--gas morphology for all four runs of a selection of clouds. Top left panels show control runs, top right ionization--only runs, bottom left winds--only runs, and bottom right winds--and--ionization runs. In general, whether or not winds acting alone were able to strongly influence the morphology of the clouds, the effects of photoionization clearly dominate at late times in the dual--feedback calculations. The relative effects of winds are strongest at early times when the massive stars are still embedded in very dense gas. Once this gas has been cleared away, the expanding HIIRs assume control of the dynamical and morphological evolution of the clouds.\\
\begin{figure*}
     \centering
  \subfloat[Run I]{\includegraphics[width=0.50\textwidth]{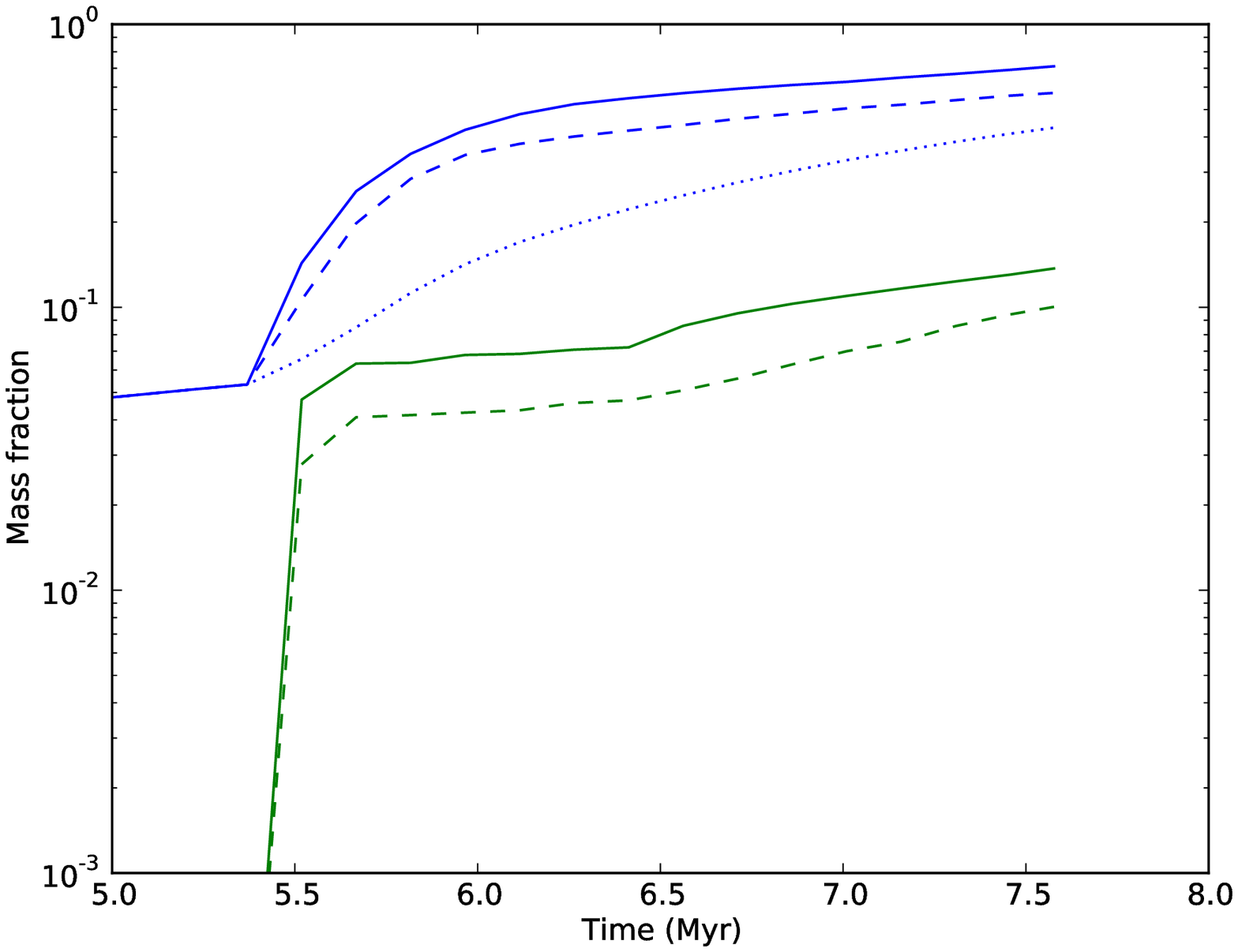}}
   \hspace{-0.05in}
      \subfloat[Run UQ]{\includegraphics[width=0.50\textwidth]{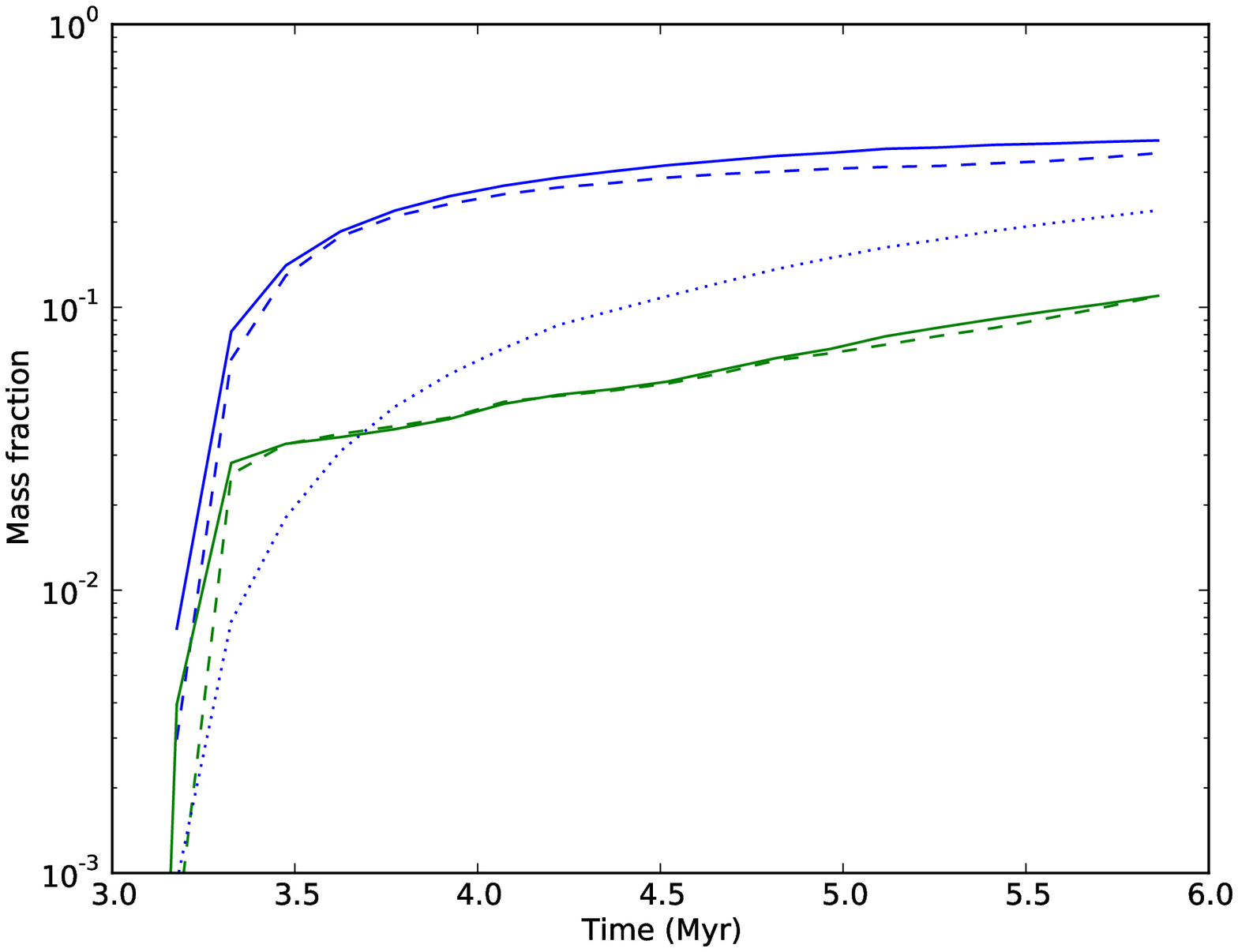}}
\caption{Evolution of the ionized (green lines) and unbound (blue lines) gas fractions in the ionization--only (dashed lines), winds--only (dotted lines) and ionization--and--winds (solid lines) Run I (left) and Run UQ (right)}
\label{fig:unbnd_iuq}
\end{figure*}
\subsection{Dynamical influence of combined winds and HII regions}
\begin{figure*}
     \centering
    \subfloat[Bound clouds]{\includegraphics[width=0.50\textwidth]{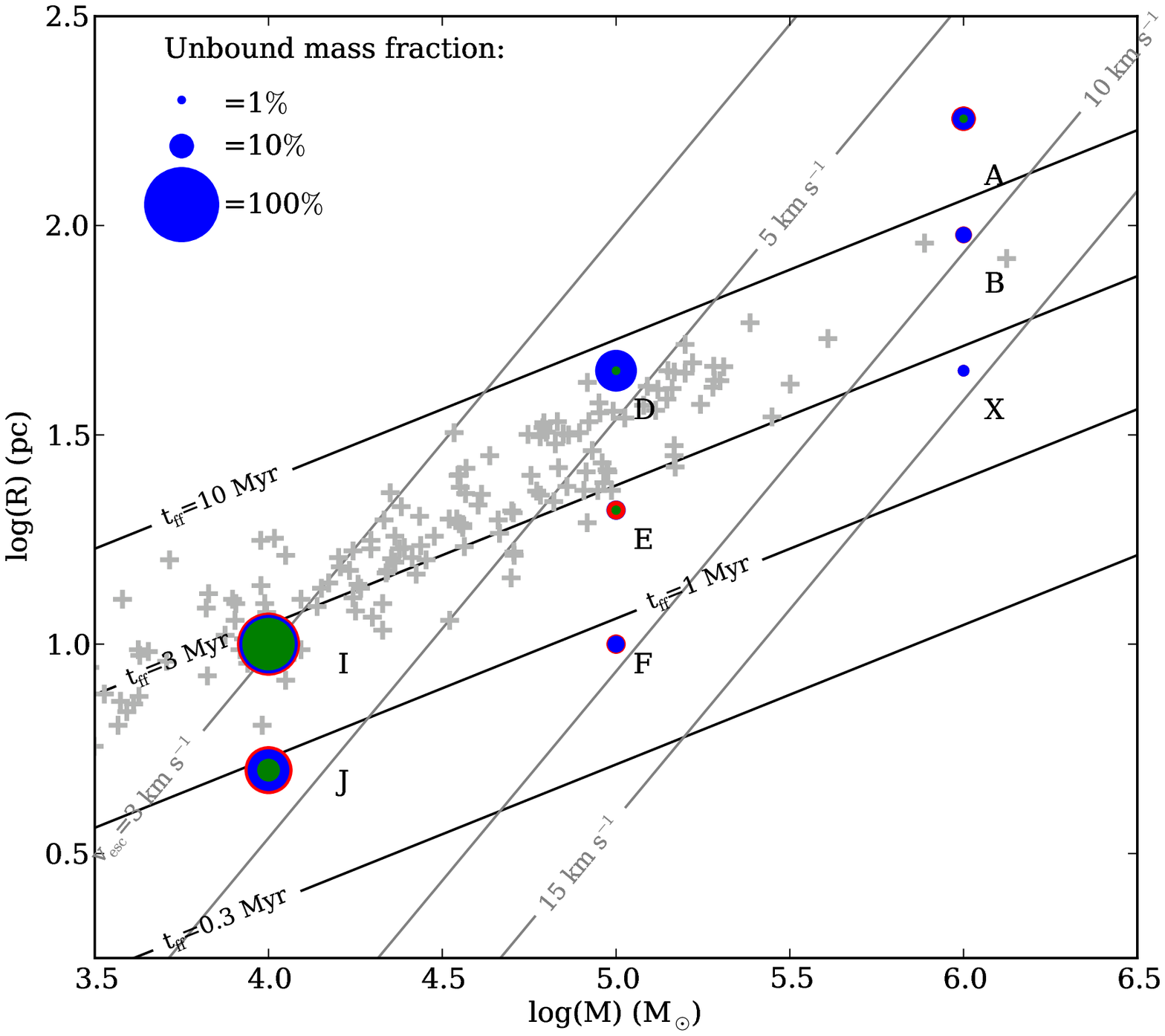}}
    \hspace{-0.05in}
    \subfloat[Unbound clouds]{\includegraphics[width=0.50\textwidth]{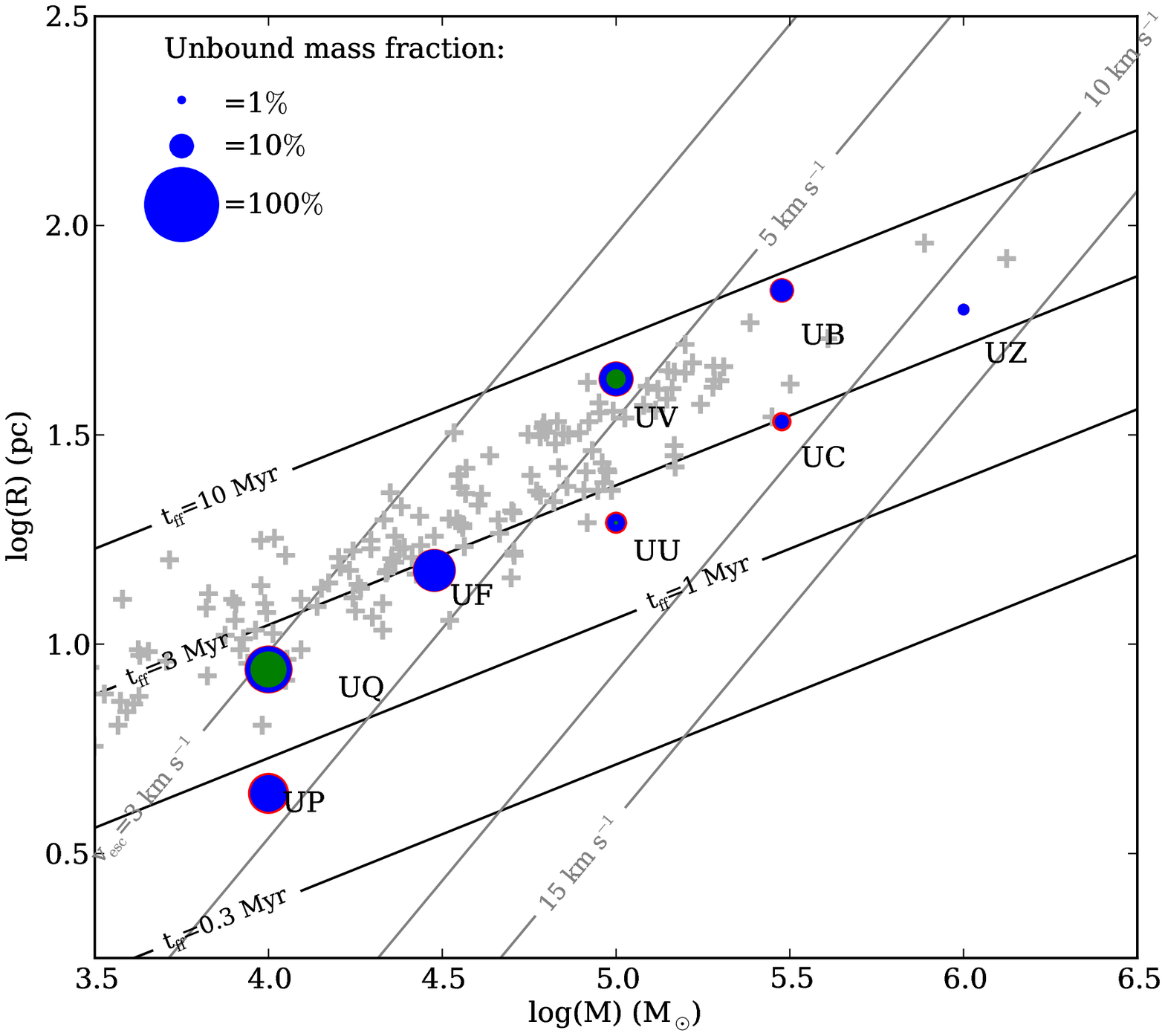}}
\caption{Fractions of system mass unbound at the ends of the bound--cloud (left) and unbound--cloud (right) simulations, denoted by blue (ionization--only), green (winds--only) and red (dual--feedback) circles.}
\label{fig:results_bnd_unbnd}
\end{figure*}

\indent The ability of stellar feedback to expel gas from GMCs is a key issue. We compute at each timestep the fraction of mass in each cloud which has positive total energy in the cloud centre--of--mass frame. Overall, we find that the combined effect of winds and ionization is to unbind substantially more mass than winds acting alone, and slightly more than ionization acting alone. Photoionization is a much more destructive agent than winds. Figure \ref{fig:unbnd_iuq} shows the evolution of the unbound mass fraction (as a fraction of the \emph{total} system mass, blue lines) and the ionized gas fraction (as a fraction of the total instantaneous gas mass, green lines) in the ionization--only (dashed lines), winds--only (dotted lines) and dual--feedback (solid lines) Runs I and UQ, these being the bound and unbound clouds with the greatest differences between ionization--only and dual--feedback runs.\\
\indent The ionization fractions are also generally slightly increased by the action of winds. Where the ionization fraction is noticeably different, the increase occurs early in the simulation, after which the rates of gas ionisation in the two calculations are very similar. This is consistent with the picture that winds are effective at early times in clearing away dense gas near the massive stars. However, once this material is destroyed, ionising photons can penetrate deeper into the gas and the dynamics of the clouds become dominated by photoionization.\\
\indent In Figures \ref{fig:results_bnd_unbnd} we show the final unbound mass mass fractions in all bound and unbound clouds as blue (ionization--only), green (winds--only) and red (dual--feedback) circles plotted over the mass--radius parameter space studied here. We include on the plot the clouds from \cite{2009ApJ...699.1092H} as light grey crosses, contours of cloud escape velocity as dark grey lines and of constant freefall time (and thus volume density) as black lines. There is a clear gradient pointing from high to low escape velocities in the ability of feedback to disrupt clouds. In particular, only clouds whose escape velocity is less than 5 km s$^{-1}$ are significantly damaged, in the sense of having more than 10$\%$ of their mass expelled over the 3Myr time window.\\
\subsection{Star formation rates and efficiencies}
\indent Feedback also affects the star formation process inside clouds. The are various ways in which this may be assessed and this has unfortunately led to some rather confusing terminology. In particular, we prefer \emph{efficiencies} to be dimensionless quantities. For clarity, we here lay out the terms we have used explicitly and explain how we have computed them from our simulations:\\
\\
(i) The \emph{absolute instantaneous star formation rate}, [SFR$(t)$]. This is defined as
\begin{eqnarray}
{\rm SFR}(t)=\frac{{\rm d}M_{*}}{{\rm d}t}.
\end{eqnarray}
We will approximate the quantity as
\begin{eqnarray}
\langle SFR(t)\rangle=\frac{M_{*}(t)-M_{*}(t_{0})}{(t-t_{0})},
\end{eqnarray}
where $t_{0}$ is the time when star formation begins, and we use units of M$_{\odot}$Myr$^{-1}$.\\
\\
(ii) The \emph{absolute star formation efficiency}, [SFE$(t)$]. We use this in the sense that it usually used in star formation and molecular cloud observations and simulations: the fraction of the total mass of system which is stellar at a given time, i.e.
\begin{eqnarray}
{\rm SFE}(t)=\frac{1}{M_{*}(t)+M_{\rm gas}(t)}\int_{t_0}^{t}{\rm SFR}(t'){\rm d}t'=\frac{M_{*}(t)}{M_{*}(t)+M_{\rm gas}(t)},
\end{eqnarray}
where $M_{*}(t)$ and $M_{\rm gas}(t)$ are respectively the instantaneous stellar and gas masses.\\
\indent \cite{2000ApJ...545..364M} defined an instantaneous star formation efficiency using the rate d$M_{*}$d$t$ at which gas is being converted to stars and the rate d$M_{\rm ej}$d$t$ at which gas is being ejected from the system in question, as $\epsilon={\rm d}M_{*}/({\rm d}M_{*}+{\rm d}M_{\rm ej})$. If the star formation rates and mass ejection rates do not vary too much over the lifetime of the system, the instantaneous and final star formation efficiencies are nearly the same. However, $\epsilon$ is only a typical value of the absolute star formation efficiency SFE$(t)$ if both quantities are small. For this reason, and the fact that $\epsilon$ does not take into account gas which is unbound at the outset of star formation, as in our unbound clouds, we concentrate on the observational quantity SFE$(t)$ here.\\
\\
(iii) The \emph{star formation efficiency rate}, [SFER$(t)$]. This is the rate of the change of the star formation efficiency, 
\begin{eqnarray}
%{\rm SFER}(t)=\frac{{\rm d}}{{\rm d}t}\left(\frac{M_{*}}{M_{*}(t)+M_{\rm gas}(t)}\right),
{\rm SFER}(t)=\frac{{\rm d}M_{*}}{{\rm d}t}\frac{1}{M_{*}(t)+M_{\rm gas}(t)},
\end{eqnarray}
which we approximate by
\begin{eqnarray}
{\rm SFER}(t)=\frac{[{\rm SFE}(t)-{\rm SFE}(t_{0})]}{(t-t_{0})}
\end{eqnarray}
This is what is often measured by galactic and extragalactic studies \citep[e.g][]{2008AJ....136.2782L}, and is unfortunately often referred to in this context as `the star formation efficiency'.\\
\\
(iv) The {star formation efficiency rate per freefall time}, [SFER$_{\rm ff}(t)$]. This denotes the change in the star formation efficiency in one freefall time \citep[e.g.][]{2007ApJ...654..304K,2011ApJ...729..133M}, and is equivalent to the star formation rate measured in absolute units multiplied by the local freefall time; 
\begin{eqnarray}
%{\rm SFER}_{\rm ff}(t)=\frac{{\rm d}}{{\rm d}t}\left(\frac{M_{*}}{M_{*}(t)+M_{\rm gas}(t)}\right)t_{\rm ff},
{\rm SFER}(t)=\frac{{\rm d}M_{*}}{{\rm d}t}\frac{t_{\rm ff}}{M_{*}(t)+M_{\rm gas}(t)},
\end{eqnarray}
and is computed by us as
\begin{eqnarray}
{\rm SFER}_{\rm ff}(t)=\frac{[{\rm SFE}(t)-{\rm SFE(t_{0})]t_{\rm ff}}}{(t-t_{0})}
\end{eqnarray}
This quantity is dimensionless.\\
\\
\indent Since all four quantities listed above may be used to infer different things about star formation and different metrics are used by different communities, we plot and discuss them in turn.\\
\subsubsection{SFE(t)}
\begin{figure*}
     \centering
    \subfloat[Bound clouds]{\includegraphics[width=0.50\textwidth]{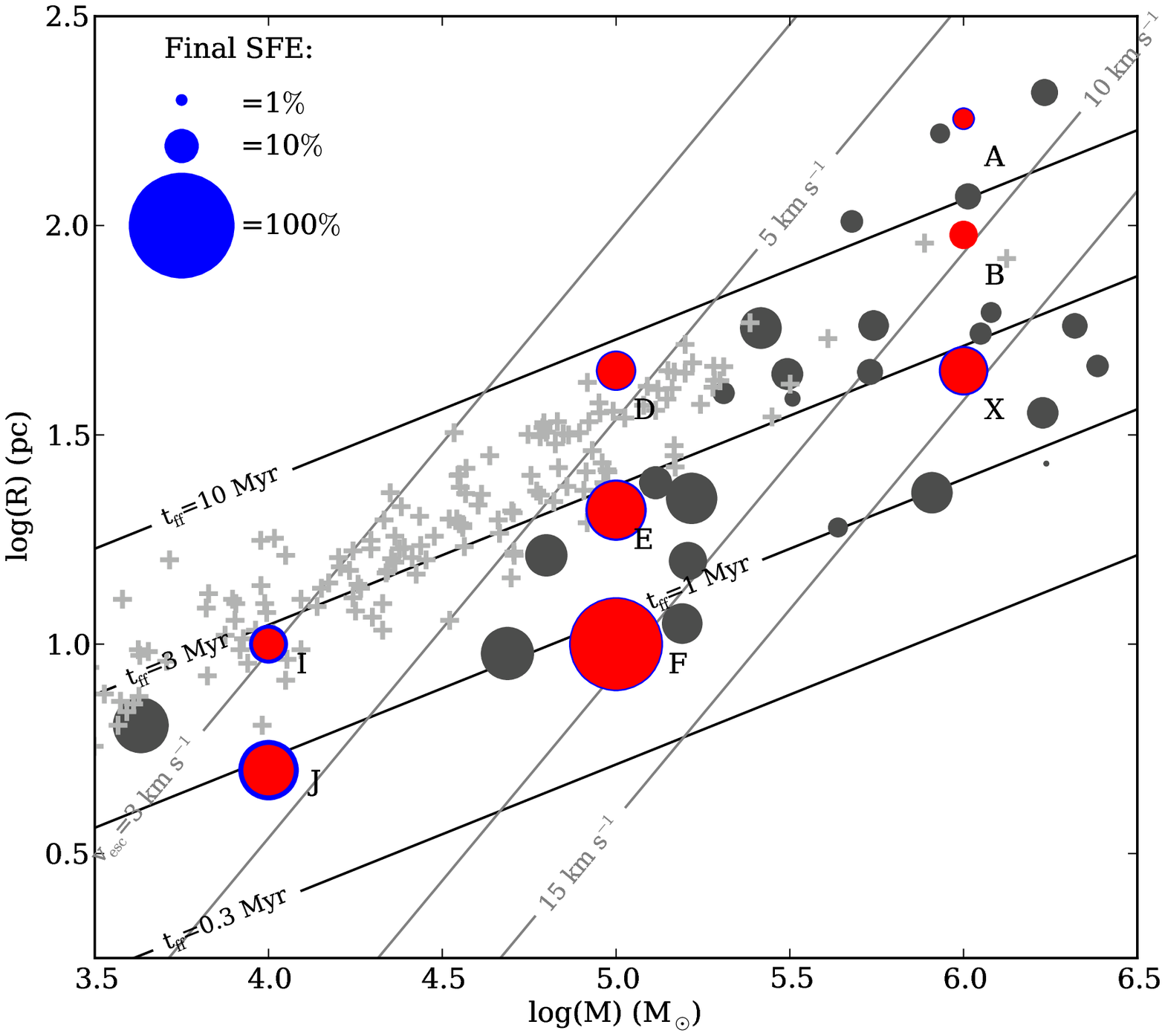}}
    \hspace{-0.05in}
    \subfloat[Unbound clouds]{\includegraphics[width=0.50\textwidth]{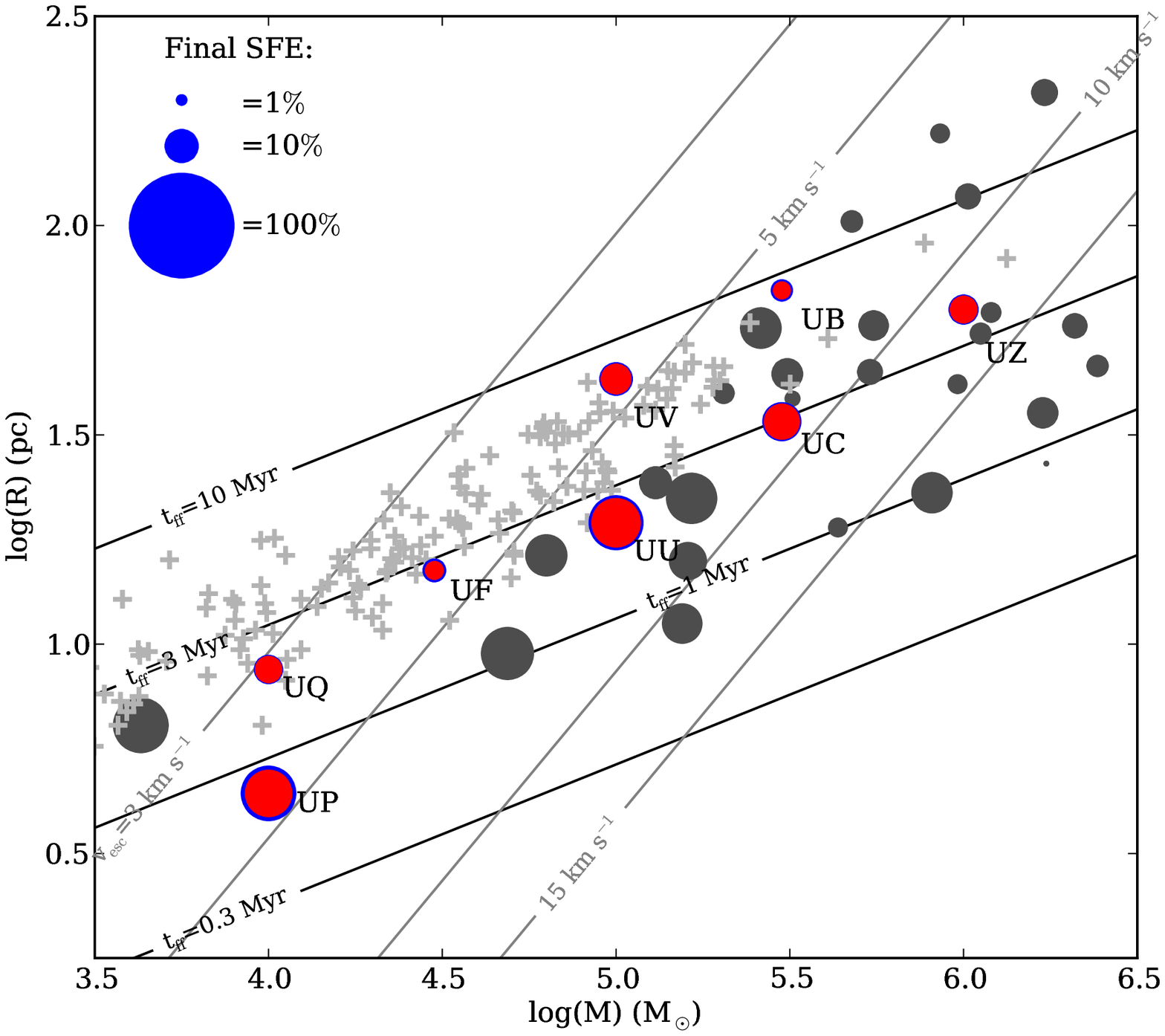}}
\caption{SFE at the ends of the bound--cloud (left) and unbound--cloud (right) simulations, denoted by blue (control runs) and red (dual feedback runs) circles. Results from Murray, 2011 are shown as grey circles.}
\label{fig:results_sfe}
\end{figure*}

\indent Figure \ref{fig:results_sfe} shows the final SFE$(t_{\rm SN})$ for all runs ($t_{\rm SN}=3$Myr, the time before the first supernova), comparing the control and dual feedback calculations to assess the ability of feedback to alter the fraction of gas converted to stars over the duration of the simulations. Coloured circles represent the simulation results and grey circles are taken from Table 2 in \cite{2011ApJ...729..133M}, who investigated 32 GMCs hosting the most luminous star--forming regions in the Milky Way.\\
\indent The initially--unbound control clouds have systematically lower values of SFE$(t_{\rm SN})$, as also found by \cite{2005MNRAS.359..809C}. The effect of feedback on the final stellar masses, denoted by the areas of the red circles relative to the blue, is modest and always negative. The effect tends to be larger for lower--mass and lower--escape velocity clouds, like the unbound mass fraction. The mean SFE for the bound clouds is reduced from by 11$\%$ from 0.233 to 0.208, and for the unbound clouds by 22$\%$ from 0.125 to 0.098.\\
\indent \cite{2009ApJS..181..321E} recently measured the SFE of five nearby clouds and find values ranging from 3--6$\%$. The \emph{total} mass of the five clouds is only $\approx10^{4}$M$_{\odot}$, equivalent to one of our runs I, J, UQ or UP, which have final values of SFE$(t)$ of 10--20$\%$, larger by a factor of around three.\\
\indent \cite{2011ApJ...729..133M} investigated clouds with masses in the range 4.3$\times10^{4}$-$8.4\times10^{6}$M$_{\odot}$ which exhibit values of SFE in the range 0.002--0.273 with a mean of 0.08. They observe a trend of decreasing SFE with increasing cloud mass, but conclude that is likely to be a selection effect. We include their results for comparison and we see that they actually agree rather well.\\
\subsubsection{SFR(t)}
\begin{figure*}
     \centering
    \subfloat[Bound clouds]{\includegraphics[width=0.50\textwidth]{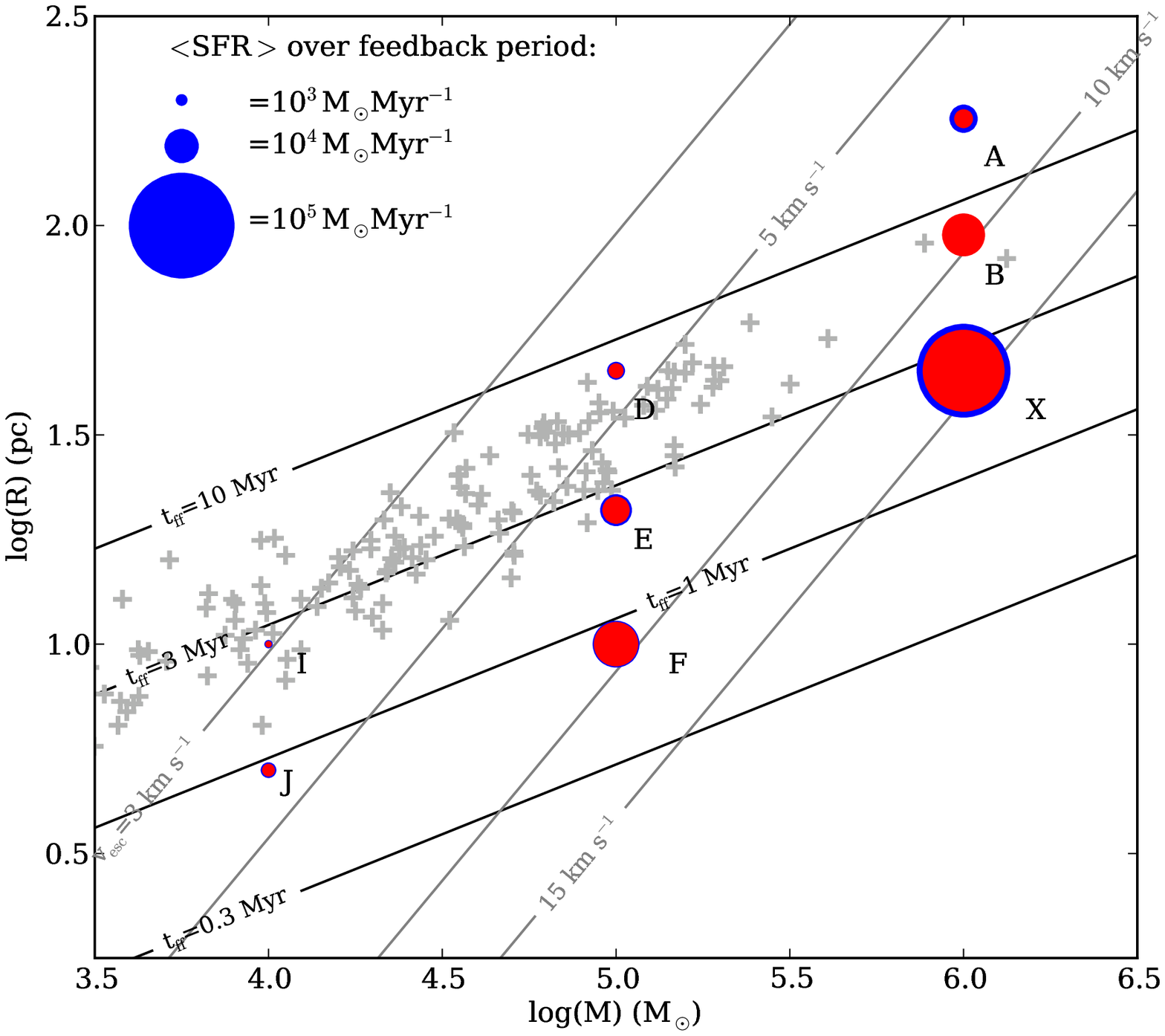}}
    \hspace{-0.05in}
    \subfloat[Unbound clouds]{\includegraphics[width=0.50\textwidth]{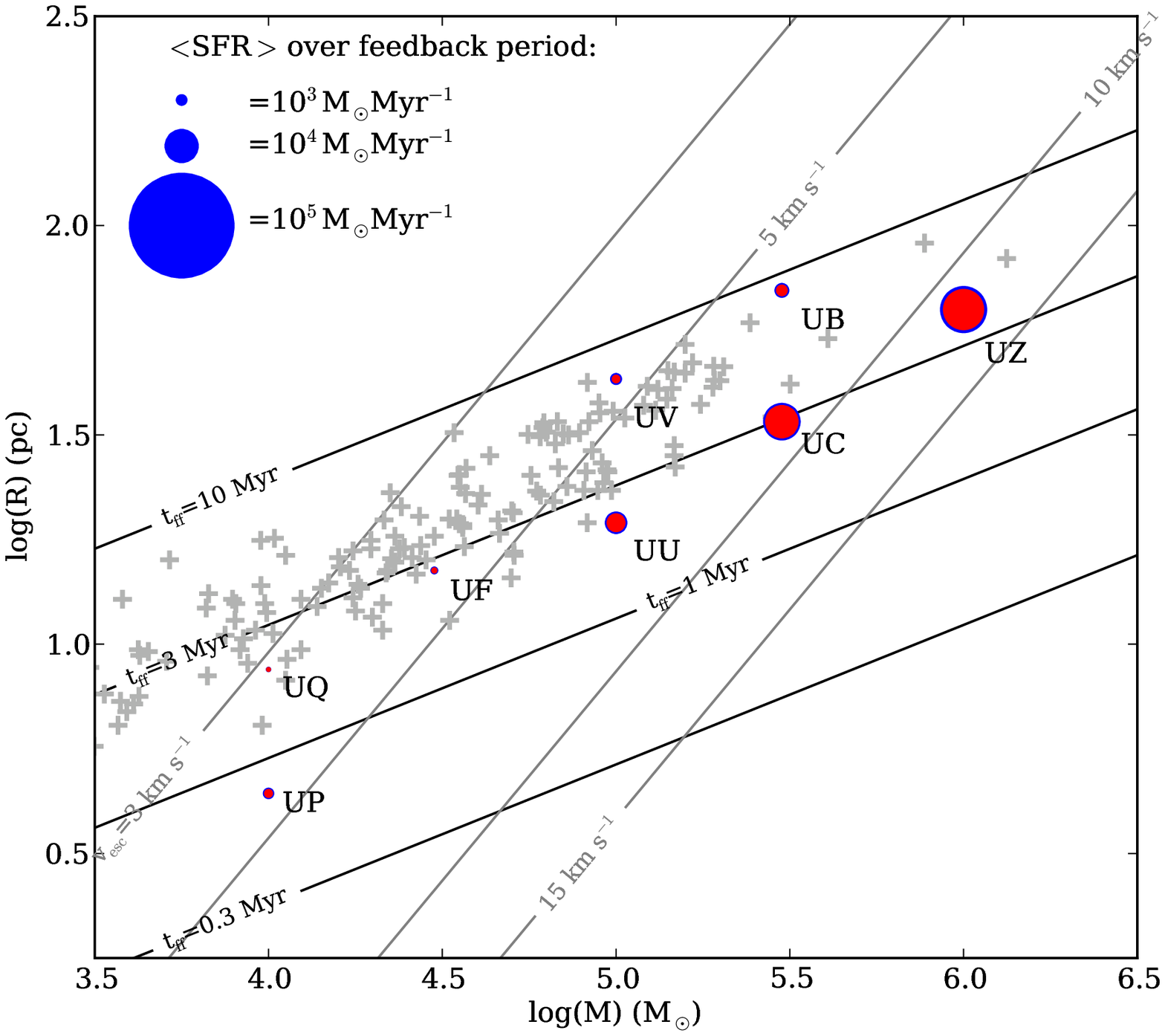}}
\caption{Mean values of SFR over the feedback interval for the bound--cloud (left) and unbound--cloud (right) simulations, denoted by blue (control runs) and red (dual feedback runs) circles.}
\label{fig:results_sfr}
\end{figure*}

\indent In Figure \ref{fig:results_sfr}, we plot the mean star formation rates over the t$_{\rm SN}$ feedback interval for all runs in units of M$_{\odot}$Myr$^{-1}$. This quantity increases with increasing cloud mass, decreases with cloud freefall time and is systematically lower in the initially--unbound clouds by factors of a few between clouds of the same mass and size. SFR$(t)$ is also only modestly influenced by feedback. Gravity remains the primary driver of star formation in these simulations.\\
\indent In their $\approx10^{4}$M$_{\odot}$ total--mass sample of nearby molecular clouds, \cite{2009ApJS..181..321E} measure a total SFR$(t)$ of 256M$_{\odot}$Myr$^{-1}$. This rate is actually very close to that seen in runs I, UQ, and UF which perhaps suggests that the observed clouds may achieve comparable SFE$(t)$ to these simulations in the future. \cite{2009ApJS..181..321E} remark that the star formation rates in their sample would result in SFE$(t)$ reaching values of 15--30$\%$ if they persisted for another 10Myr.\\
\subsubsection{SFER(t)}
\begin{figure*}
     \centering
    \subfloat[Bound clouds]{\includegraphics[width=0.50\textwidth]{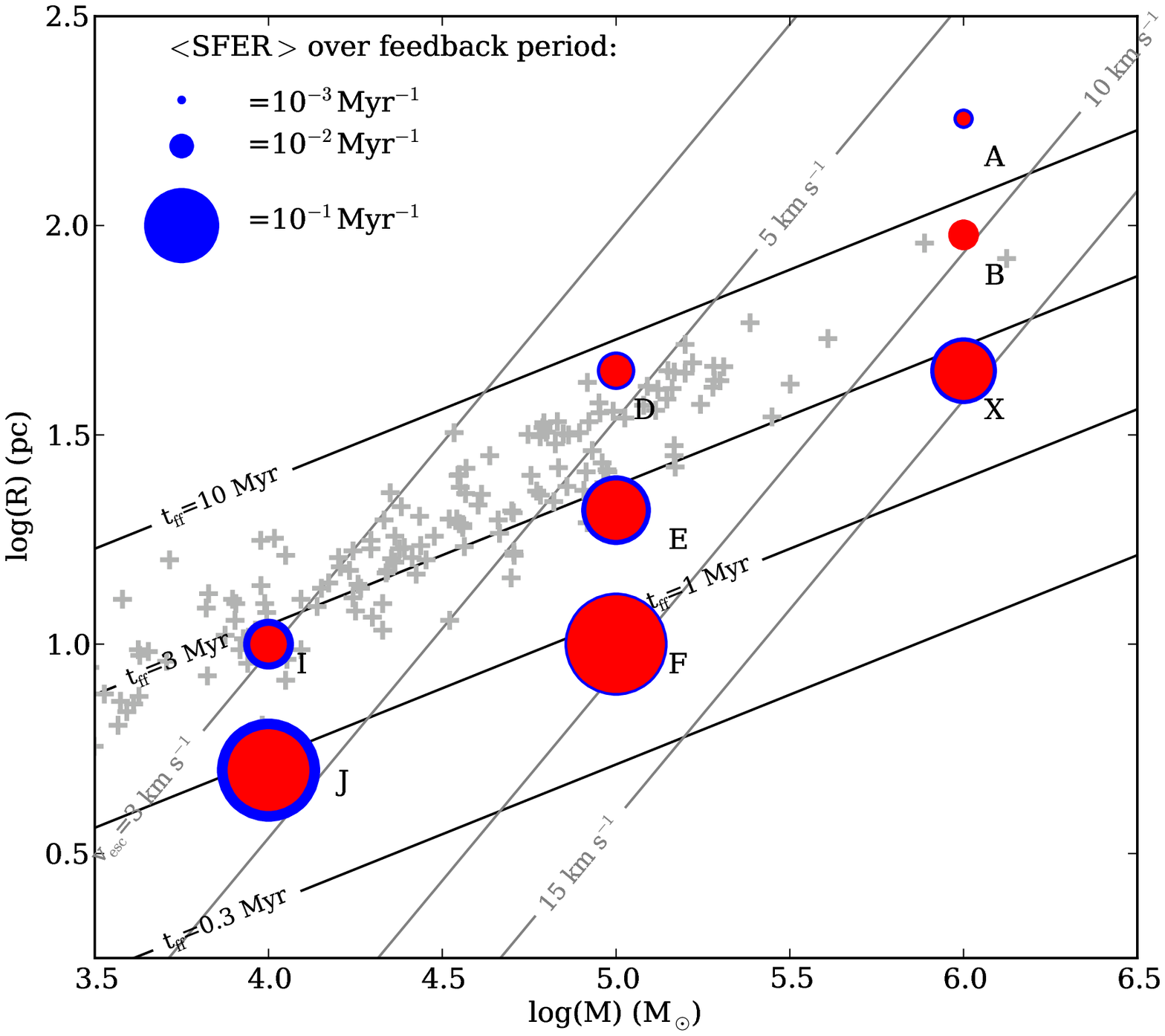}}
    \hspace{-0.05in}
    \subfloat[Unbound clouds]{\includegraphics[width=0.50\textwidth]{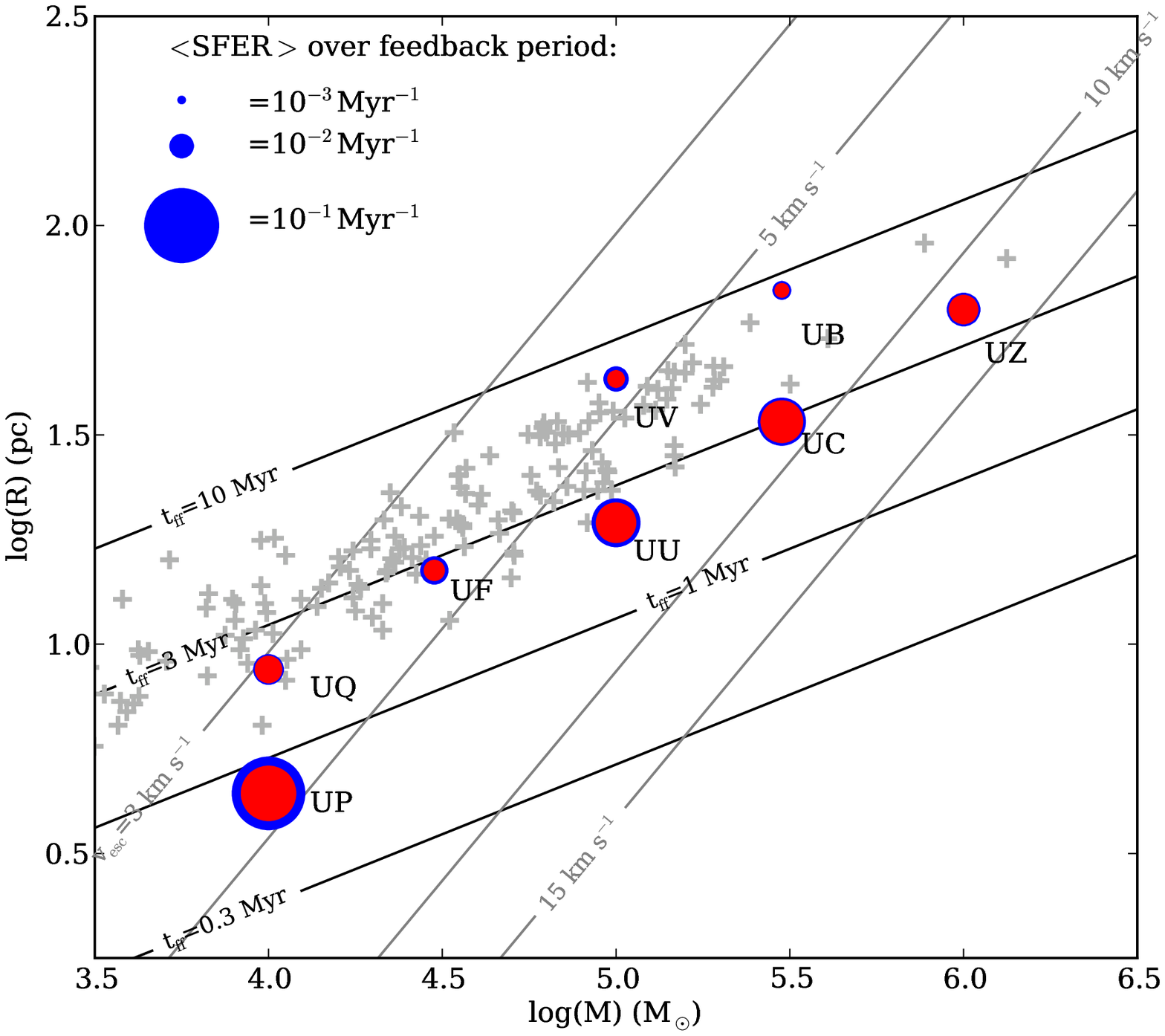}}
\caption{Mean values of SFER$(t)$ over the feedback interval for the bound--cloud (left) and unbound--cloud (right) simulations, denoted by blue (control runs) and red (dual feedback runs) circles.}
\label{fig:results_sfer}
\end{figure*}

\indent We plot the star formation efficiency rate SFER$(t)$ in units of Myr$^{-1}$ in Figure \ref{fig:results_sfer}. This quantity varies little with cloud mass but increases substantially with decreasing cloud freefall time. SFER$(t)$ is considerably smaller in the initially unbound clouds, having typical values of a few percent Myr$^{-1}$, whereas it reaches $\approx 10\%$Myr$^{-1}$ for bound clouds with freefall times of 1Myr. Feedback clearly reduces SFER$(t)$ significantly for several of the smaller model clouds.\\
\indent It is instructive to compare these values with the survey of SFER$(t)$ in 23 nearby galaxies in \cite{2008AJ....136.2782L}. They compute SFER$(t)$ as
\begin{eqnarray}
{\rm SFER}(t)_{\rm obs}=\frac{\Sigma_{\rm SFR}}{\Sigma_{\rm gas}},
\end{eqnarray}
where $\Sigma_{\rm SFR}$ is the star formation rate surface density in M$_{\odot}$Myr$^{-1}$pc$^{-2}$, and $\Sigma_{\rm gas}$ is the gas surface density in M$_{\odot}$pc$^{-2}$. The typical gas surface densities for the molecule--dominated regions of the spirals in their sample are a few$\times10$M$_{\odot}$pc$^{-2}$ and for the dwarfs an order of magnitude less.\\
\indent They obtain typical values of SFER$(t)$ in regions of spiral galaxies where the gas mass is largely molecular of $\sim5\times10^{-4}$Myr$^{-1}$ (they obtain a similar typical value for dwarf galaxies where the ISM mass is dominated by HI). This quantity is $\sim10^{2}$ times smaller than the values of SFER$(t)$ we compute here. However, the observed values are likely to be decreased by the dilution of star--forming gas in the telescope beams with non--star--forming material, raising $\Sigma_{\rm gas}$ relative to $\Sigma_{\rm SFR}$.\\
\indent \cite{2009ApJS..181..321E} discuss their results in light of this issue. For their sample of low--mass clouds, they obtain SFER$(t)$ of 3--6$\%$ over the last 2Myr, corresponding to $1.5-3\%$Myr$^{-1}$, comparable to the lower range of values we measure. However, we note that their define the borders of their clouds as the $A_{\rm V}=2$ contour, corresponding to $\approx6\times10^{-3}$g cm$^{-3}$. Using the same criterion for our own clouds would reduce by factors of a few to ten the quantities of non--star--forming gas and thus increase the apparent SFE by the same factor. We therefore conclude that our clouds are forming stars a factor of at least a few too quickly.\\
\begin{figure*}
     \centering
    \subfloat[Bound clouds]{\includegraphics[width=0.50\textwidth]{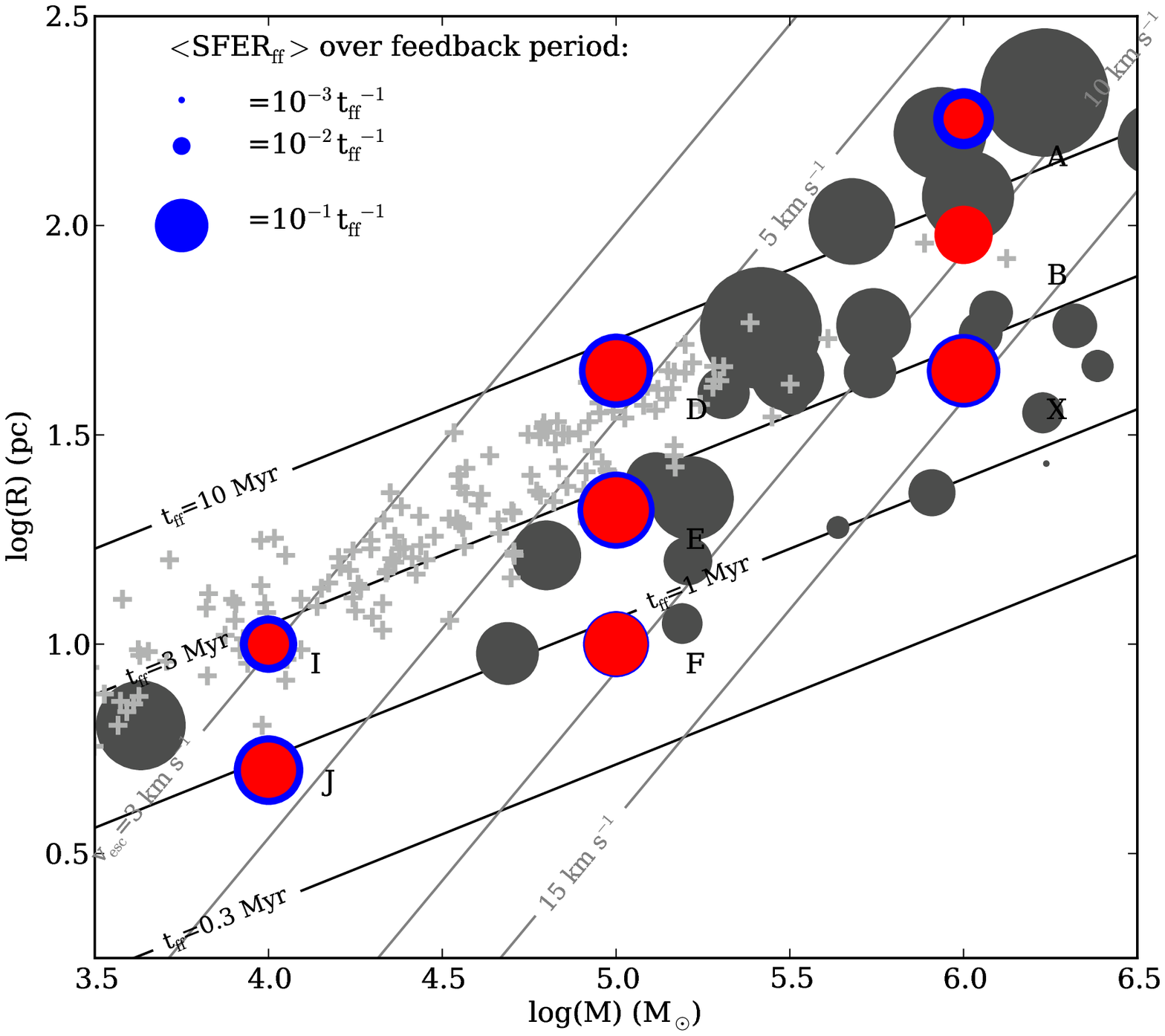}}
    \hspace{-0.05in}
    \subfloat[Unbound clouds]{\includegraphics[width=0.50\textwidth]{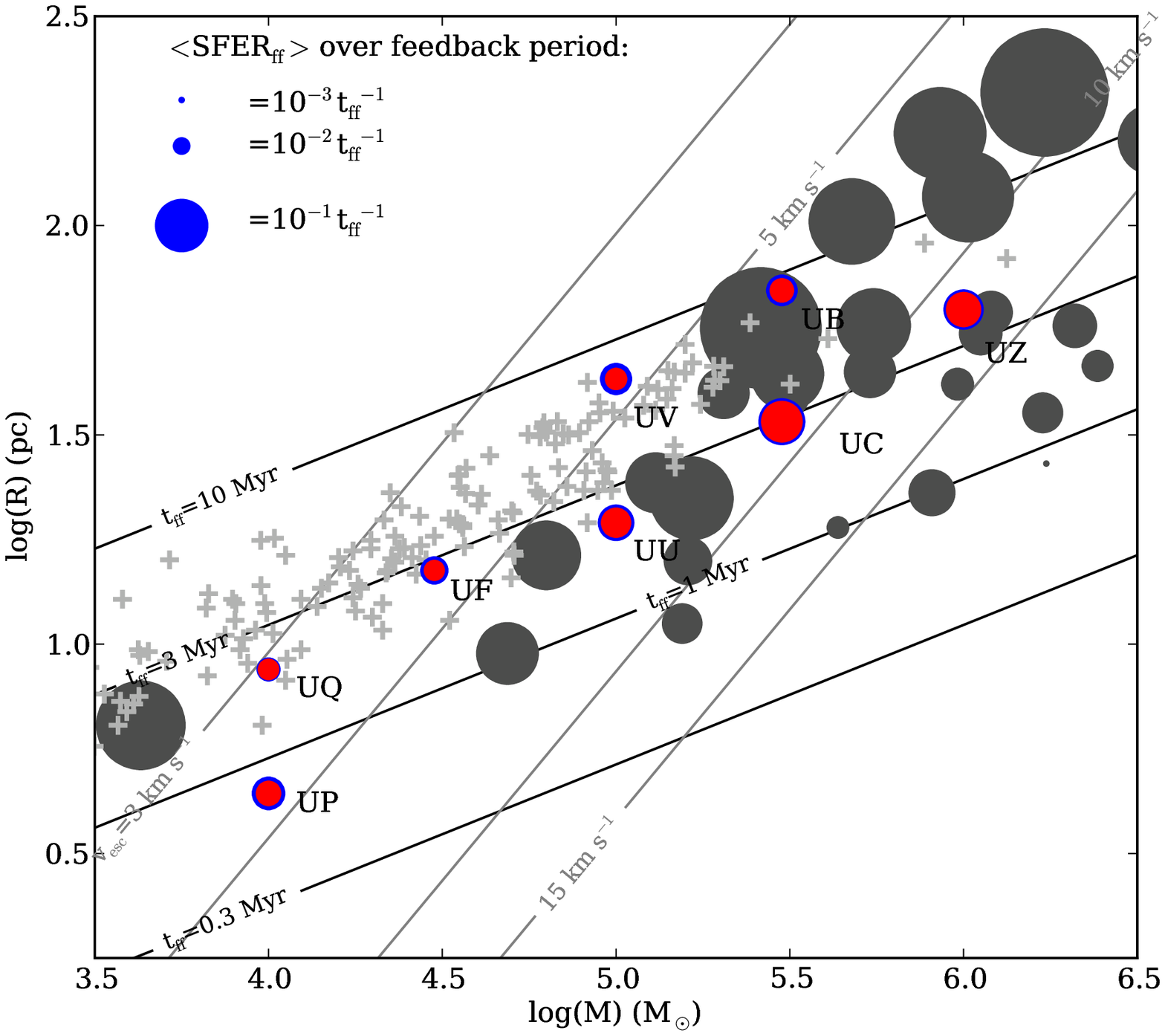}}
\caption{Mean values of SFER$_{\rm ff}(t)$ over the feedback interval for the bound--cloud (left) and unbound--cloud (right) simulations, denoted by blue (control runs) and red (dual feedback runs) circles. Results from Murray, 2011 are shown as grey circles.}
\label{fig:results_sferff}
\end{figure*}
\subsubsection{SFER$_{\rm ff}$(t)}
\indent Finally, in Figure \ref{fig:results_sferff}, we plot SFER$_{\rm ff}(t)$, again including the data from \cite{2011ApJ...729..133M}. This plot shows that, in general, the initially--unbound clouds convert gas to stars more slowly and \emph{the cloud boundedness is the dominant parameter determining} SFER$_{\rm ff}$, although feedback does have some influence. SFER$_{\rm ff}$ is nearly constant in the bound clouds at values of order 10$\%$, and nearly constant in the bound clouds at values of around $3\%$.\\
\indent \cite{2007ApJ...654..304K} measured SFER$_{\rm ff}(t)$ in a variety of systems of different densities and inferred typical values of $\approx0.02$, essentially independent of density (although see \cite{2007ApJ...668.1064E} for further discussion of these measurements). \cite{2009ApJS..181..321E} obtain values of SFER$_{\rm ff}(t)$ in the range 0.03--0.06, for entire clouds, but rather larger values (0.1--0.25) for dense cores. The values of SFER$_{\rm ff}(t)$ computed by \cite{2011ApJ...729..133M} are generally larger than in our unbound clouds, but are comparable in the case of the bound clouds. Their observed values are in the range 0.001--0.592 with a mean of 0.16.\\
\indent In our bound clouds, the mean SFER$_{\rm ff}(t)$ is reduced by feedback by 29$\%$ from 0.160 to 0.113 and in the unbound clouds by 32$\%$ from 0.038 to 0.026. The unbound clouds have intrinsically lower values, which was also  observed in simulations of unbound clouds by \cite{2005MNRAS.359..809C}. The effect of feedback is relatively slight and overall, the bound clouds form stars a few times faster than suggested by \cite{2009ApJS..181..321E} and 5--8 times faster than the systems quoted in \cite{2007ApJ...654..304K}, but at comparable rates to those observed by \cite{2011ApJ...729..133M}. The unbound clouds are towards the lower limits of these observational estimates.\\
\subsection{HII region photon leakage}
\begin{figure*}
     \centering
    \subfloat[Bound clouds]{\includegraphics[width=0.50\textwidth]{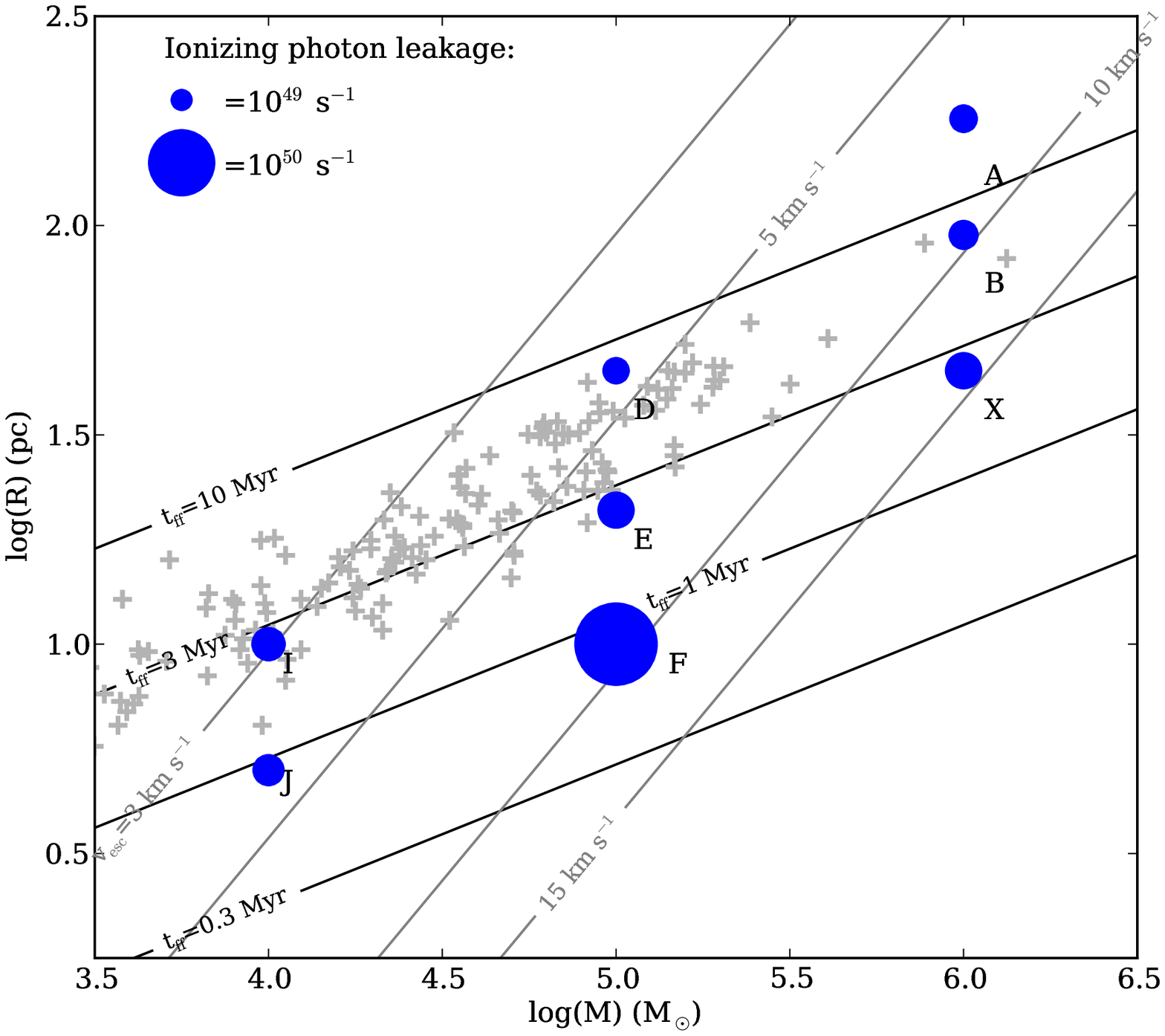}}
    \hspace{-0.05in}
    \subfloat[Unbound clouds]{\includegraphics[width=0.50\textwidth]{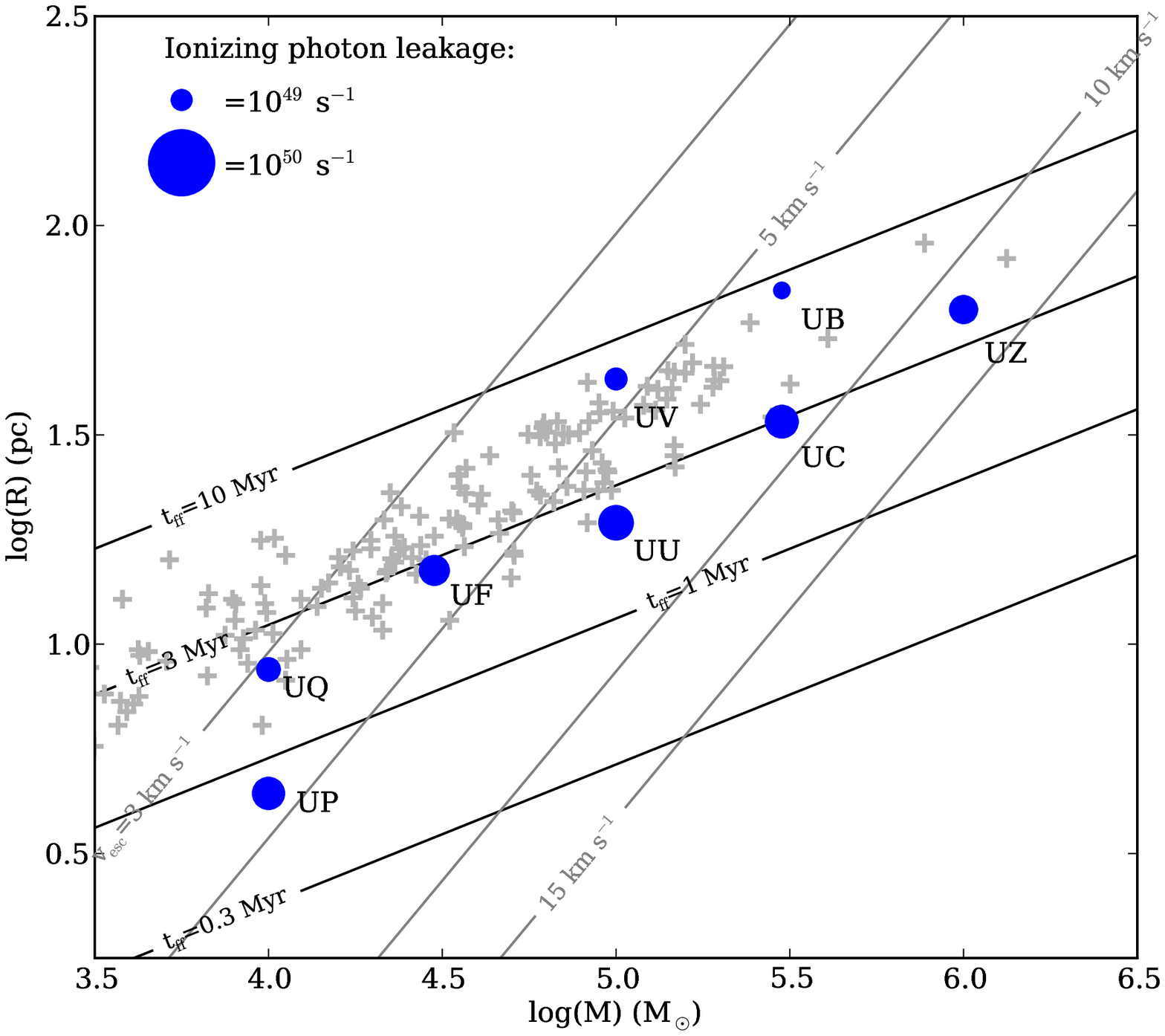}}
\caption{Effective Lyman continuum luminosities for the bound--cloud (left) and unbound--cloud (right) simulations, denoted by red (control runs) and blue (dual feedback runs) circles.}
\label{fig:results_phot}
\end{figure*}

We compute the fluxes of ionizing photons leaking from the clouds. Comparing these results to those from Papers I and III shows modest differences, with the additional action of winds making the clouds slightly more leaky to photons, particularly in the very early stages of the evolution. The lower--mass clouds in particular lose substantial fractions (0.5-0.9) of their ionizing photons. A similar degree of photon loss was reported by \cite{2012MNRAS.427..625W} in their studies of the disruption of 10$^{4}$M$_{\odot}$ fractal clouds.\\
\indent In Figure \ref{fig:results_phot}, we plot effective ionizing luminosities for each cloud. Although the clouds' actual fluxes and photon leakage fractions vary considerably, they do so in opposite directions, so that the effective fluxes (with the exception of the strongly gas--depleted Run F) fall in a narrow range. The more massive and denser clouds tend to have higher absolute luminosities because of their higher star formation efficiencies. These are also the clouds least affected by feedback and therefore least likely to lose photons. There is a clear trend for the bound clouds to be more luminous, in absolute and effective terms, than the unbound clouds because of their higher SFE's. In general, all clouds (save Run F) have an effective luminosity close to 10$^{49}$s$^{-1}$. Run F is bright because of its very high star formation efficiency and it being close to gas exhaustion by the end of the feedback time window. Such behaviour is expected of starburst clouds.\\
\subsection{Transparency to supernova ejecta}
\begin{figure*}
     \centering
    \subfloat[Bound clouds]{\includegraphics[width=0.50\textwidth]{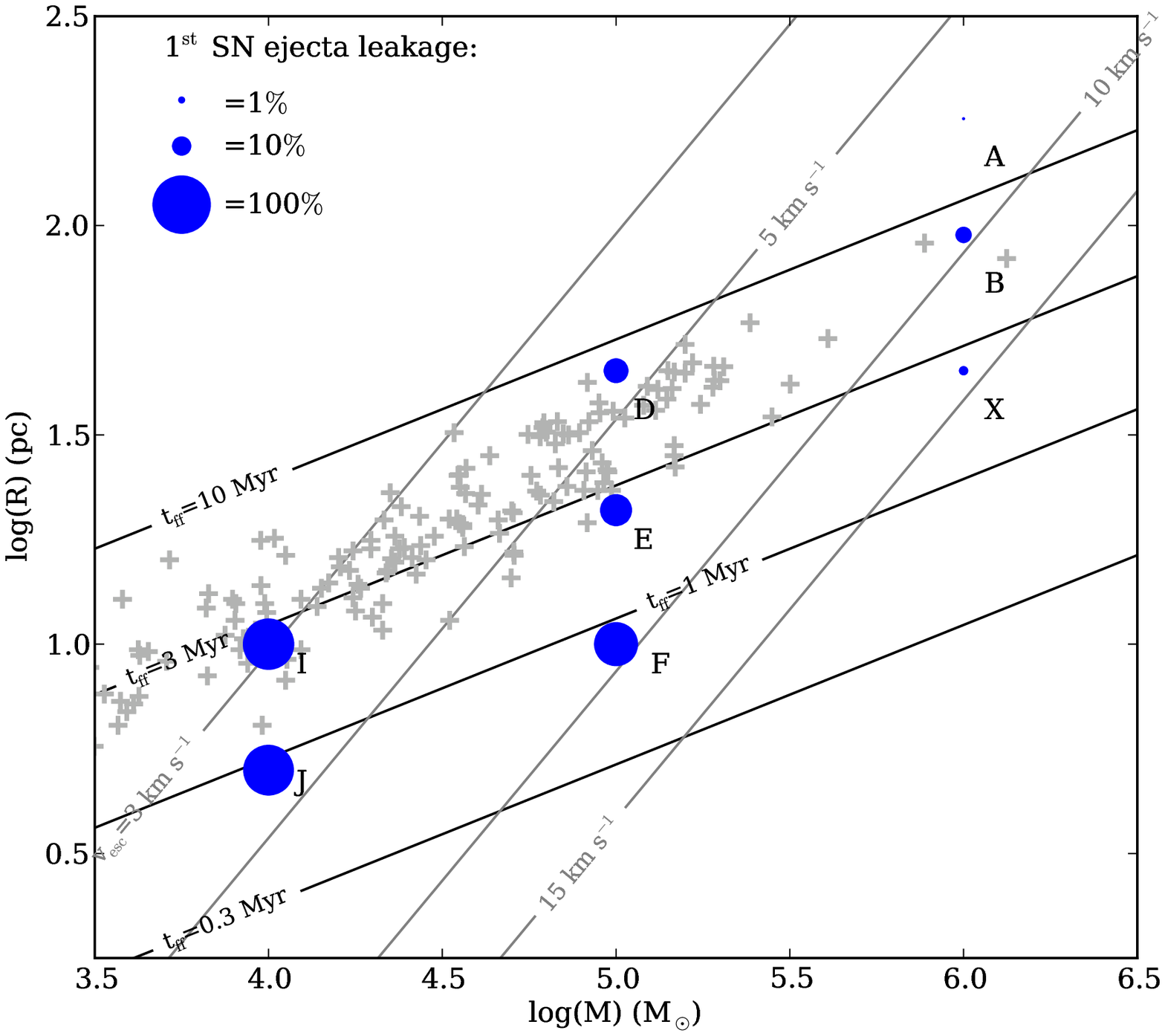}}
    \hspace{-0.05in}
    \subfloat[Unbound clouds]{\includegraphics[width=0.50\textwidth]{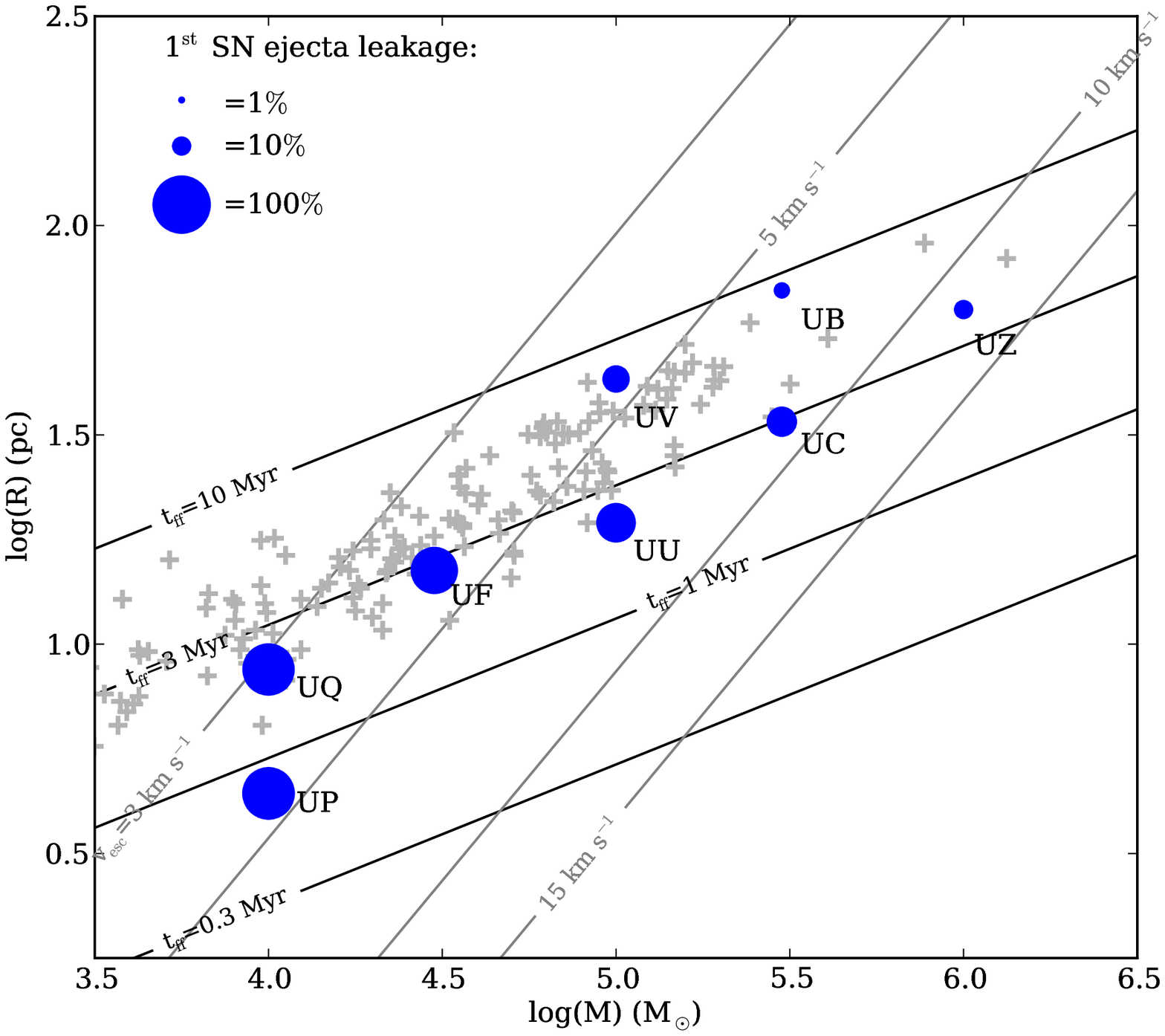}}
\caption{Fraction of ejecta lost from first supernovae for the bound--cloud (left) and unbound--cloud (right) simulations, denoted by red (control runs) and blue (dual feedback runs) circles.}
\label{fig:results_ejecta}
\end{figure*}
\indent Figure \ref{fig:results_ejecta} depicts the fraction of the ejecta from the first supernova explosion in each cloud which is expected to escape. This was computed in the same fashion as in Papers I and III.\\
\indent All but the most massive clouds will leak substantial fractions of their first supernova ejecta. Conversely, while a few of the lowest--mass clouds -- Run I, J and UQ -- will be essentially destroyed by their first SN, large quantities of material in the more massive and denser clouds are likely to survive. It is therefore possible that there will be a second round of star formation involving debris--polluted gas in these clouds. This is especially true of the more massive clouds which will likely be able to withstand several supernova explosions, as is the case in the 30 Doradus region \citep[][and references therein]{2006AJ....131.2140T}.\\
\section{Discussion}
\subsection{Effects of HII regions versus winds}
\begin{figure}
\includegraphics[width=0.45\textwidth]{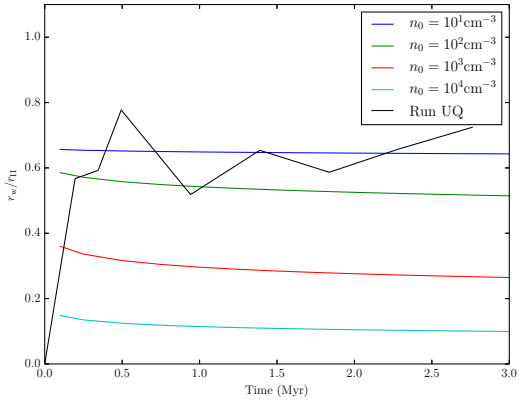}
\caption{Evolution of the ratio of the wind bubble radius to the HII region radius as a function of time for four values of $n_{0}$ as computed from Equation 15 (coloured lines) and as measured in Run UQ (black line).}
\label{fig:rwrii}
\end{figure}

Our treatment of stellar winds as purely sources of momentum is plainly simplified and limits the range of phenomena that we can observe. Since we do not inject hot wind gas into our calculations, we do not correctly model interactions between the wind and the molecular material such as Kelvin--Holmholtz instabilities or thermal/hydrodynamical ablation of cold material by the hot wind. This limits the ability of our winds to entrain material from the clouds\\
\indent \cite{2013MNRAS.431.1337R} simulated the effects of the winds and supernovae (but not the radiation) of three central O--stars on turbulent molecular clumps with masses $\approx3\times10^{3}$ and $\approx1\times10^{4}$M$_{\odot}$, the latter being similar to our Runs I, J, UP and UQ. They observed that the hot wind gas was easily able to escape the clouds via low--density channels and that it was able to efficiently entrain mass on the way, achieving mass--loading factors of $10^{2}-10^{3}$. However, the typical mass ejection rates (for the larger cloud) were a few $\times10^{-4}$M$_{\odot}$ yr$^{-1}$, comparable to the rates at which these clouds are destroyed by winds alone in Paper V, but roughly an order of magnitude slower than the rates at which material is unbound from our $10^{4}$M$_{\odot}$ clouds by photoionisation. Direct comparison of calculations including and neglecting different physics is difficult and dangerous, but this suggests that our conclusion that ionisation is the main driver of cloud destruction is likely to be correct.\\
\indent By injecting momentum alone, we have implicitly assumed that the shocked stellar wind is able to cool maximally effectively. The thermodynamics of wind bubbles interacting with HII regions are in reality more complex. This issue was recently studied in one--dimensional simulations by \cite{2014arXiv1403.1620M}. There is always an initial period during which the wind remains hot and the pressure in the wind bubble within the HII region is very high. This drives a shock into the ionised gas, sweeping up an ionized shell. If this shell becomes sufficiently dense, recombinations within it may consume all of the O--stars' ionising flux, trapping the HII region inside the wind bubble. However, the duration of this phase depends strongly on the ability of the wind to cool. There are two major mechanisms which are likely to control when this occurs: evaporation from the inner wall of the ionised shell and mass loading from ablation/evaporation of dense cold clumps inside the wind bubble.\\
\indent The former was considered by \cite{1988ApJ...324..776M} who computed a cooling time for the bubble (their Equation 14). This timescale is short ($\sim10^{5}$yr) for most of our simulations, although $\sim$1Myr for a few of our largest, lowest--density clouds such as Run A, if they are treated as a single bubble and not as several interacting bubbles.\\
\indent There are numerous small cold clumps inside our HII regions and wind bubbles and photevaporation of these objects would further mass load a real wind and shorten its cooling time further. We compared their evaporation timescales in the simulations to those given by \cite{2007dmsf.book..245P} and references therein and find good agreement, indicating that these and other photoevaporative flows in the simulations are adequately resolved. In common with \cite{2013MNRAS.431.1337R}, we find that these objects are rather long--lived, often surviving for $\sim1$Myr. Their contribution to mass loading would be small compared to thermal evaporation at the edge of the wind bubble, as estimated by \cite{1988ApJ...324..776M}.\\
\indent We conclude that momentum--dominated winds are a reasonable assumption for the embedded clusters modelled here and observe that they are a rather small perturbation on the effects of the HII regions. Our findings that the winds are confined by the HII regions are in agreement with those of several other authors \citep[e.g][]{1984ApJ...278L.115M,2009ApJ...703.1352K,2012ApJ...757..108Y}.\\
\indent We alluded briefly to this issue in Paper V where we simply compared in Figure 9 the expansion laws of momentum--driven wind bubbles and HII regions and showed that, except at very early times or extreme densities, the HII region was always larger. We here give a slightly more sophisticated analysis. We consider a wind bubble expanding into an HII region. The HII region has radius $r_{\rm II}(t)$, density $\rho_{\rm II}(t)$, pressure $P_{\rm II}(t)$ and internal sound speed $c_{\rm II}=10$km s$^{-1}$. The wind bubble has radius $r_{\rm w}(t)<r_{\rm II}(t)$. The driving star has an ionising photon flux $Q_{\rm H}=10^{49}$s$^{-1}$, a wind mass loss rate $\dot{M}=10^{-6}$M$_{\odot}$yr$^{-1}$ and a wind terminal velocity $v_{\infty}=2000$km s$^{-1}$, and the original neutral medium has a mass density $\rho_{0}$ and a number density $n_{0}$. At time $t$, the radius of the unperturbed HII region is given by
\begin{eqnarray}
r_{\rm II}(t)=R_{s}\left(1+\frac{7}{4}\frac{c_{\rm II}t}{R_{\rm s}}\right)^{\frac{4}{7}}
\end{eqnarray}
\citep{1978ppim.book.....S} and the pressure by
\begin{eqnarray}
P_{\rm II}(t)=\rho_{\rm II}(t)c_{\rm II}^{2}=\rho_{0}\left(1+\frac{7}{4}\frac{c_{\rm II}t}{R_{\rm s}}\right)^{-\frac{6}{7}}c_{\rm II}^{2}
\end{eqnarray}
\citep{2009A&A...497..649B}, where the pressure is assumed to be uniform, and $R_{\rm s}$ is the Str\"omgren radius,
\begin{eqnarray}
R_{\rm s}=\left(\frac{3Q_{\rm H}}{4\pi n_{0}^{2}\alpha_{\rm B}}\right)^{\frac{1}{3}}.
\end{eqnarray}
\indent We now make the simple assumption that a wind bubble expands inside the HII region until the ram pressure at $r_{\rm w}$ is equal to the pressure inside the HII region, which is modified from its unperturbed value $P_{\rm II}$ to a new value $P_{\rm II,w}$ as
\begin{eqnarray}
P_{\rm II,w}=P_{\rm II}\frac{r_{\rm II}^{3}}{(r_{\rm II}-r_{\rm w})^{3}}.
\end{eqnarray}
The effects of the winds are assumed not to alter the radius of the HII region or the ionised gas mass. We may then equate the pressures and, after some rearrangement, we arrive at
\begin{eqnarray}
\begin{array}{rl}
\frac{1}{r_{\rm w}^{2}}=&\frac{r_{\rm II}^{3}}{(r_{\rm II}-r_{\rm w})^{3}}\left(\frac{4\pi\rho_{0}c_{\rm II}^{2}}{\dot{M}v_{\infty}}\right)\left(1+\frac{7}{4}\frac{c_{\rm II}t}{R_{\rm s}}\right)^{-\frac{6}{7}}\\
&\\
=&\frac{r_{\rm II}^{3}}{(r_{\rm II}-r_{\rm w})^{3}}\beta(t).
\end{array}
\end{eqnarray}
If we assume that $r_{\rm}<r_{\rm II}$, we are justified in taking only the first two terms in the expansion of $(r_{\rm II}-r_{\rm w})^{3}$ and we obtain an expression for $r_{\rm w}$:
\begin{eqnarray}
r_{\rm w}^{2}+\frac{3r_{\rm w}}{\beta(t)r_{II}(t)}-\frac{1}{\beta(t)}=0.
\end{eqnarray}
\indent We plot the evolution of this function for several values of $n_{0}$ in Figure \ref{fig:rwrii}. Since the initial pressure of the HII region scales with $n_{0}$, the ratio $r_{\rm w}/r_{\rm II}$ increases with decreasing $n_{0}$. However, the wind bubble is always substantially smaller than the HII region. \cite{2012ApJ...757..108Y} define a parameter $\Omega=P_{\rm w}r_{\rm w}^{3}/(P_{\rm II}r_{\rm II}^{3}-P_{\rm w}r_{\rm w}^{3})$ to quantify the importance of winds in HII regions. The models presented above have $\Omega$ in the range 0.1--1 The influence of radiation pressure (discussed below) is likely to be small, so these models fall into the class of classical Str\"omgren HII regions (see Figure 1 in \cite{2012ApJ...757..108Y}).\\
\indent The above analysis applies to a spherical non--leaking bubble. In a leaky bubble, the pressure inside the HII region drops much faster than it would otherwise, which allows the wind bubble to expand to occupy a greater proportion of the HII region interior. We also plot in Figure \ref{fig:rwrii} the time evolution of $r_{\rm w}/r_{\rm II}$ measured in Run UQ. This calculation's initial number density is $\approx 10^{3}$cm$^{-3}$, but the evolution of $r_{\rm w}/r_{\rm II}$ is closer to that of a cloud with n$_{0}=10$cm$^{-3}$.\\
\indent The two principal influences of photoionization on the cold gas -- ionisation of fresh material and the thermal pressure of the HIIR on the inner walls of the bubbles -- are not much influenced by the winds in our calculations. Although the ionized gas morphology and behaviour are somewhat affected, the differences in the behaviour of the cold dense gas are modest. The photoevaporation flows at the ionisation front effectively protect the molecular material from the wind, and the evolution of the cold gas is largely controlled by photoionization.\\
\indent \cite{2013MNRAS.435...30W} measured linewidths in the ionized material on the outer surfaces of two gaseous pillars in NGC 3603. While greater than the ionized sound speed, the inferred gas velocities were much lower than the free wind velocities of typical O--stars, or than the expected sound speed in hot shocked wind gas. They concluded that the winds were not interacting with the pillars directly, but with the photoevoration flows being driven from their surfaces, in agreement with what we observe here.\\
\subsection{Global cloud evolution}
\indent The influence of feedback on the global evolution of our model clouds was generally modest. The quantities of gas actually ionized are small and do not vary a great deal amongst calculations, generally being 3--10$\%$. Typical photoevaporation rates are $\sim$10$^{-2}$M$_{\odot}$yr$^{-1}$ for the 10$^{6}$M$_{\odot}$ clouds and $\sim$ few$\times$10$^{-3}$M$_{\odot}$yr$^{-1}$ for the 10$^{5}$M$_{\odot}$ and 10$^{4}$M$_{\odot}$ clouds, so do not vary very much with cloud mass. They are lower in clouds with larger escape velocities for a given mass.\\
\indent Several authors have examined this problem analytically \citep[e.g][]{1979MNRAS.186...59W,1997ApJ...476..166W}. \cite{2002ApJ...566..302M} derived the following expression relating the quantity of matter photoevaporated, $\delta M_{\rm dest}$ to the ionisation timescale $t$, cloud column--density $N_{\rm H_{2}}$ and mass $M_{\rm c}$, and the total ionising luminosity $Q_{\rm H}$:
\begin{eqnarray}
\begin{array}{ll}
\delta M_{\rm dest}=&1.2\times10^{4}M_{\odot}\left(\frac{t}{3.7{\rm Myr}}\right)^{\frac{9}{7}}\left(\frac{N_{\rm H_{2}}}{1.5\times10^{22}{\rm cm}^{-2}}\right)^{-\frac{3}{14}}\times\\
&\left(\frac{M_{\rm c}}{10^{6}{\rm M}_{\odot}}\right)^{\frac{1}{14}}\left(\frac{Q_{\rm H}}{10^{49}{\rm s}^{-1}}\right)^{\frac{4}{7}}.
\end{array}
\label{eqn:matzner}
\end{eqnarray}
This expression is strictly only valid for blister HII regions and the analysis which generated it ignores both gravity and cloud internal structure. However, comparing  with the quantities of gas photoionized in our simulations (to be strictly consistent, we compared with our ionization--only calculations, but these differ little from the dual--feedback simulations presented here), we find rather good agreement, usually within a factor of two. Equation \ref{eqn:matzner} predicts less photoevaporation for clouds with low and intermediate escape velocities (the largest discrepancy being by a factor of 4.5 for Run D), and more for clouds with high escape velocities, the largest discrepancy being by a factor of 2.6 for Run X. As well as being in being in reasonable consensus, the trend in the discrepancies points in the expected direction in that the clouds with the highest escape velocities, where gravity plays a greater role, are those for which Equation \ref{eqn:matzner} overestimates the ionised mass.\\
\indent \cite{2006ApJ...653..361K} and \cite{2011ApJ...738..101G} perform one--dimensional simulations of the effects of HII regions on clouds over much longer timescales than we are able to address here.  \cite{2006ApJ...653..361K} find that 50--70$\%$ of their 2$\times10^{5}$ and 1$\times10^{6}$M$_{\odot}$ clouds are evaporated on timescales of 10 and 20 Myr respectively, which implies an average of 7--10$\%$ per 3Myr. These rates are larger than we observe but again not by very large factors, at most around five. \cite{2011ApJ...738..101G} find average photoevaporation rates in their accreting--cloud models of order 10$^{-2}$ M$_{\odot}$ yr$^{-1}$, comparable to our 10$^{6}$M$_{\odot}$ clouds but factors of 3--5 larger than our 10$^{5}$M$_{\odot}$ clouds. These calculations involve gravity but do not account for photon leakage or inhomogeneities in the gas structure, which may explain why we observe smaller photoevaporated masses. In general, our results agree with the above--cited works that the destruction timescales for large GMCs can be long. Given that we expect such clouds to also survive several supernovae, our results also allow cloud lifetimes of 10-20 Myr.\\
\indent Only the lower--density and lower--mass clouds had substantial fractions of their mass unbound over $t_{\rm SN}$. The amount of material unbound is controlled by the cloud escape velocity relative to the fixed ionised sound speed.\\
\indent Changes in the star formation rates and efficiencies were negative but slight -- never as much as a factor of two, even for the low--mass clouds most severely damaged by feedback. Clouds formed stars on approximately their freefall timescales. Both of these results suggest that feedback is largely unable to wrest control of the clouds from gravity.\\
\indent \cite{2010ApJ...715.1302V} find that their HII--region like feedback is not very effective in \emph{destroying} their model clouds, although they found that it could be destructive on smaller scales. This is in general agreement with our results -- feedback is able to destroy small-- and medium--scale structures, but generally struggles to dismantle the largest clouds. These findings are broadly in agreement with those of \cite{2010ApJ...715.1302V} and \cite{2012MNRAS.427..625W}, who also find that photionization is damaging to lower--mass ($\sim10^{4}$M$_{\odot}$ clouds). Semianalytic models by \cite{2006ApJ...653..361K} also concluded that lower--mass clouds should survive for shorter times than the largest clouds.\\
\indent In Table \ref{tab:rates}, we compare the star formation efficiencies, rates, efficiency rates and efficiency rates per freefall time in our dual--feedback models with those of \cite{2007ApJ...654..304K}, \cite{2009ApJS..181..321E} and \cite{2011ApJ...729..133M}. Our model clouds form stars too fast and too efficiently by most measures by factors of a few. Feedback has only a modest influence on the rate and efficiency of star formation.\\
\begin{table*}
\begin{tabular}{|l|l|l|l|}
Quantity & Simulations & Observations & Reference\\
\hline
SFE & 0.05--0.20 & 0.03--0.06 & \cite{2009ApJS..181..321E}\\
&&0.08 & \cite{2011ApJ...729..133M}\\
\hline
SFR (M$_{\odot}$Myr$^{-1}$) & 150--350 (Runs I, UF, UQ) & 256 & \cite{2009ApJS..181..321E}\\
\hline
SFER (Myr$^{-1}$) &0.03--0.10 & 0.015--0.030 & \cite{2009ApJS..181..321E}\\
\hline
SFER$_{\rm ff}$ & 0.03--0.1 & 0.02 & \cite{2007ApJ...654..304K}\\
&&0.03--0.06 &\cite{2009ApJS..181..321E}\\
&&0.16 &  \cite{2011ApJ...729..133M}\\
\end{tabular}
\caption{Comparison of the various measures of star formation rates and efficiencies from Section 3 between our dual--feedback simulations and observations.}
\label{tab:rates}
\end{table*}
\indent This results contrast somewhat with those of \cite{2010ApJ...715.1302V}. They studied the effects of stellar feedback on flattened clouds molecular clouds formed by colliding flows. They found that, depending on the size scales of substructure in the clouds, the SFE was reduced by factors of 3--10. They use a slightly different definition of SFE from us; SFE$=M_{*}/(M_{\rm dense}+M_{*})$, where $M_{\rm dense}$ is the mass of gas whose density exceeds 50 cm$^{-3}$. They observe that feedback \emph{increases} the quantity of dense gas while \emph{decreasing} the stellar mass. This, combined with the definition of SFE given above, makes the amplitude of variation of the SFE larger than what we measure here, using the \emph{total} gas mass, although not by large factors. It is not clear to us why feedback has a substantially greater influence on star formation in \cite{2010ApJ...715.1302V}'s calculations, although it may be related to the flattened geometry of their clouds. We will discuss this issue in greater detail in a companion paper.\\
\indent There is an additional reason that the final stellar masses of our denser clouds tend to be less affected by feedback. Since we have assumed that feedback is driven purely by O--stars and all our calculations start from starless clouds, each cloud must first form O--stars before feedback can act. O--stars or O--star--bearing clusters take time to form and each cloud continues to form lower--mass stars/clusters in the meantime. Clouds which overall have higher fractional rates of conversion of mass to stars (i.e. higher values of SFER$(t)$) are likely to have converted more of their gas reserves to stars before forming their first O--stars and to therefore be more difficult to unbind. We again defer a detailed discussion of this issue to a later paper.\\
\subsection{Neglected physical effects}
\indent In these calculations, we have neglected some important physics, particularly:\\
\indent (i) Small--scale feedback such as jets and outflows and thermal feedback from accretion onto protostars which should reduce the local star formation efficiency by factors of a few. \cite{2000ApJ...545..364M} examined the influence of momentum input from outflows on the formation of low--mass clusters and concluded that they were able to restrain the star--formation efficiencies to values of 30--50$\%$. Thermal feedback from protostellar accretion affects the fragmentation process \citep[e.g][]{2005ApJ...628..817M,2007MNRAS.382L..30S}. \cite{2009MNRAS.398...33P} modelled this process on the scale of a small cluster. They found that it limited fragmentation on small scales, reducing the numbers of stars formed and the total stellar mass by factors of up to a few. This issue can also be approached by asking whether multiple outflows drive turbulence, which supports clouds against collapse on large scales. \cite{2006ApJ...640L.187L,2007ApJ...662..395N} and \cite{2007ApJ...659.1394M} all concluded that outflows can be efficient drivers of turbulence. \cite{2010ApJ...709...27W} modelled the influence of magnetic fields and outflows on the star formation rates in turbulent cores. They showed that they have complementary effects, together decreasing SFRs by a factor of approximately three over timescales of $\sim 1t_{\rm ff}$.\\
\indent (ii) Magnetic fields, which are likely to depress the overall rate of star formation on large and small scales. Numerous recent studies \citep[e.g][]{2009MNRAS.398...33P, 2010ApJ...709...27W,2011MNRAS.414.2511V,2011ApJ...730...40P,2011ApJ...729...72P,2012ApJ...761..156F,2014arXiv1401.6096M} have found that the support provided at intermediate and large scales by magnetic fields slows star formation rates by factors of a few up to an order of magnitude in the case of \cite{2011MNRAS.414.2511V}. The interaction of HIIRs with magnetic fields has been less intensively studied. \cite{2007ApJ...671..518K} and \cite{2011ApJ...729...72P} both find that magnetic fields constrain the expansion of HIIRs, at least in some directions. \cite{2012ApJ...745..158G} find that that the presence of magnetic fields results in a much larger injection of energy into the cold gas, although much of the additional energy is stored in the magnetic field. \cite{2011MNRAS.414.1747A} came to a similar conclusion in their simulations of magnetised HIIRs growing around O and B stars, finding that large quantities of energy were stored in the magnetic fields, but that the differences in bubble evolution were modest.\\ 
\indent (iii) Radiation pressure, which has effects similar to stellar winds but is rather more difficult to model. A star's total radiative momentum flux $\dot{p}_{\rm RAD}$ is given by $L_{\rm bol}/c$ where $L_{\rm bol}$ is the star's bolometric luminosity. The stars's wind momentum flux $\dot{p}_{\rm WIND}$ is $\dot{M}v_{\infty}$, where $\dot{M}$ is the wind mass loss rate and $v_{\infty}$ is its terminal velocity. Given expressions that describe the bolometric luminosity, mass loss rate and wind terminal velocity as functions of mass, the relative contribution of radiation pressure and winds from a given star may be computed. Here we take $L_{\rm bol}(M)$ to be given by 
\begin{eqnarray}
\begin{array}{llr}
L_{\rm bol}(M)=&(M/M_{\odot})^{4}L_{\odot}&M<2M_{\odot}\\
&0.73(M/M_{\odot})^{3.5}L_{\odot}&2M_{\odot}<M<20M_{\odot}\\
&1140(M/M_{\odot})L_{\odot}&20M_{\odot}<M,\\
\end{array}
\label{eqn:l_of_m}
\end{eqnarray}
\citep[][]{2010ApJ...710L.142F}. The functions used to compute the wind mass loss rates and terminal velocities are given in Paper V as
\begin{eqnarray}
\dot{M}(M_{*})=\left[0.3{\rm ~exp}\left(\frac{M_{*}}{28}\right)-0.3\right]\times10^{-6}{\rm M}_{\odot}{\rm yr}^{-1}
\label{eqn:mdot}
\end{eqnarray}
and
\begin{eqnarray}
v_{\infty}(M_{*})=\left[10^{3}(M_{*}-18)^{0.24}+600\right]{\rm km~s}^{-1}
\label{eqn:vinf}
\end{eqnarray}
We neglect in these calculations feedback from all stars less massive than 20M$_{\odot}$, and no star more massive than 65M$_{\odot}$ exists in any simulation. Evaluating the ratio $\dot{p}_{\rm RAD}/\dot{p}_{\rm WIND}$ for these stars gives 0.82 and 0.18 respectively, since the wind momentum declines steeply toward lower masses, whereas the radiation momentum varies linearly with mass. The stars with the largest momentum flux, which dominate the total flux, are therefore those for which radiation pressure would make the smallest relative contribution. Overall, radiation pressure is a perturbation of order unity to the wind momenta for individual stars in this mass range. This quantity can be straightforwardly integrated over the stellar mass function, adequately described by a Salpeter function between 0.5 and 65M$_{\odot}$ for simulations in which we resolve stars, and a similar function for those in which we can only resolve clusters. If we again ignore winds entirely for stars less massive than 20M$_{\odot}$ but include the radiation pressure from all stars, the total radiative momentum flux is $\approx50\%$ larger than the total wind flux. However, this still represents a correction of order unity which is unlikely to substantially influence our results.\\
\indent \cite{2014MNRAS.439.2990S} simulated the effects of direct radiation pressure and/or photoionisation on uniform and isothermal--profile clouds. They found that radiation pressure was able to accelerate large quantities of gas to high velocities, but took a long time to do so. If photoionisation was also active, it was much more rapidly--acting and tended to dominate.\\
\indent In common with \cite{2014MNRAS.439.2990S, we have neglected in the above analysis} multiple photon scatterings, which increases the coupling of the radiative momentum flux to the gas by a factor termed f$_{\rm trap}$ by \cite{2009ApJ...703.1352K}, who argue that its value should be $\approx2$. However, in protocluster clouds with large optical depths (i.e. with large surface--, but not necessarily volume--, densities), f$_{\rm trap}$ can be larger. \cite{2010ApJ...710L.142F} take f$_{\rm trap}=$2-5 and argue that radiation pressure from massive ($>10^{4}$M$_{\odot}$) clusters in dense ($\Sigma>$a few$\times10^{-2}$--$10^{-1}$g cm$^{-2}$) clouds dominates over all other feedback mechanisms. \cite{2010ApJ...709..191M} argue in a similar fashion and present models of W49 and G298.4-0.3, which have surface densities in the range 0.05--0.1 g cm$^{-2}$, showing that radiation pressure forces are always larger than HII gas pressure forces.\\
\indent While very few individual sub--clusters in our simulations reach $10^{4}$M$_{\odot}$, the total stellar mass in many of our large clouds exceeds this value and some clouds (e.g. Runs X and F) have surface densities in the regime where these authors find radiation pressure to be important. It is therefore possible that radiation pressure would be significant in some of our models, and we will address this issue in future work.
\subsection{Leakage of photons from HIIRs}
\indent The loss of ionising photons by massive star forming regions is a key ingredient in three major areas of astrophysics, namely the energetics of the large--scale ISM, the inference of star formation rates in unresolved stellar populations (for example, in dwarf galaxies), and the reionization of the universe.\\
\indent The source of photons required to maintain the diffuse ionised gas (DIG) in spiral galaxies is still a matter of debate, but there seems to be a consensus that O--stars are responsible for most of it but that the contribution of field O--stars is not sufficient, so that embedded massive stars must make up much of the difference \citep{2006ApJ...644L..29V}. This in turn requires that the clouds in which these stars are embedded are porous to photons. Whether this is the case can only be addressed by GMC--scale simulations of the kind presented here. This issue has been addressed before using, for example, ionising sources in static fractal density distributions \citep[e.g.][]{2005ApJ...633..295W} but it is only relatively recently that estimates could be self--consistently made from dynamical simulations in which the dynamics is largely driven by ionisation itself \citep[e.g][]{2012MNRAS.427..625W}.\\
\indent We found that all of our clouds, with the exception of the severely gas--depleted Run F, leak ionizing photons at close to the same rate of $\sim10^{49}$s$^{-1}$ regardless of mass. Their specific effective ionising luminosities are therefore proportional to $M_{\rm cloud}^{-1}$. This in turn implies that, for any cloud mass function where there are more low--mass clouds than high--mass clouds, such as in M33 \citep[][]{2003ApJS..149..343E}, most of the photons ionizing the ISM originate from low--mass clouds.\\
\indent The questions of inferring stellar IMFs and star formation histories from the emission of massive stars and of reionization of the universe are concerned with the issue of the escape of photons from \emph{galaxies}. \cite{2008ApJ...672..765G} for example use an AMR simulation with minimum comoving resolutions of $\approx50$pc to study the escape of ionising photons high--redshift galaxies, computing source luminosities using STARBURST99 \citep[][]{1999ApJS..123....3L}. This resolution limit is roughly the size of a whole GMC and therefore cannot include the physics in our calculations, which would modify the emission of clusters from the STARBURST models. The simulations presented here may therefore be useful as subgrid inputs for large--scale calculations.\\
\section{Conclusions}
In this and preceding papers, we have examined the evolution of a set of sixteen model GMCs characterised by radius, mass and initial turbulent velocity dispersion under four different physical assumptions: they suffer no feedback at all from their stars, that they experience ionizing radiation only, that they experience winds only, or that they experience both forms of feedback. Modulo the simplifying assumptions we have made about modelling the feedback, we are now in a position to draw some conclusions about the relative and total influence of winds and photoionization:\\
\indent (i) Except at very early times when the OB--stars are still deeply embedded, the dynamical effects of the two forms of feedback are dominated by photoionization. The additional influence of the winds on the cold gas and on the star formation process is essentially a perturbation.\\
\indent (ii) Winds are able to substantially modify the morphology of the \emph{ionized} gas in two ways. At early stages of the cloud evolution, when the HIIRs are still breaking out of the dense filamentary gas where the O--stars are located, winds help clear the gas away and reduce collimation of the ionized flows. This produces roughly spherical HIIRs more closely resembling common morphological types observed by \cite{1989ApJS...69..831W} than the multi--lobed structures seen in the ionization--only calculations.\\
\indent At later stages, the wind--blown HIIRs often exhibit central holes with typical sizes $\sim10$pc, closely corresponding to the shell--like morphological class identified by \cite{1989ApJS...69..831W}. Overall, the shapes of the windblown HIIRs are more realistic than those from the ionization--only simulations.\\ 
\indent (iii) As with ionization acting alone, the combined effects of winds and ionization on the cloud dynamics can be substantial when compared with the absence of any feedback. In the 3Myr time window before the first supernovae are expected, winds and ionization are able in some cases to unbind more than half the mass of their host clouds. However, the degree of influence is strongly constrained by the clouds' escape velocities. While feedback is very destructive to the lower--mass and lower--density clouds modelled here, it has very little effect on the more massive and denser objects.\\
\indent (iv) The influence of winds and ionization on star formation rates, efficiencies and efficiency rates is rather modest. In particular, the mean final SFEs of the bound and unbound clouds are reduced by 11$\%$ and 22$\%$ respectively to mean values of 0.1--0.2. Mean SFER$_{\rm ff}$'s are reduced by around one third in both cloud samples to values of 3--11$\%$. By most measures, the clouds form stars too fast and too efficiently in general, and feedback does relatively little to change this picture.\\
\indent (vi) The model clouds leak substantial quantities of ionizing photons. The leakiness and intrinsic luminosity anticorrelate so that the effective luminosities of the clouds are nearly independent of mass at a few$\times10^{49}$s$^{-1}$.\\
\indent (vii) Most of the clouds are also leaky with respect to supernova debris and all except the lowest--density low--mass clouds are likely to survive at least one supernova explosion.\\ 
\indent (viii) As well as photons, the HIIRs are also permeable to their own ionized gas. The fact that the HII regions are leaky strongly affects the dynamics of the bubbles and limits the amount of damage they can do to their host clouds.\\
\section{Acknowledgements}
We thank the referee, Chris Matzner, for very interesting comments and suggestions which resulted in a more comprehensive manuscript.
This research was supported by the DFG cluster of excellence `Origin and Structure of the Universe' (JED, BE).  IAB acknowledges funding from the European Research Council for the FP7 ERC advanced grant project ECOGAL. JN would like to  acknowledge the German Deutsche Forschungsgemeinschaft, DFGPR 569/9-1.

\bibliography{myrefs}

\label{lastpage}

\end{document}